\documentclass[aps,prc,twocolumn,floatfix,superscriptaddress]{revtex4}
\usepackage{longtable}
\usepackage{graphicx}
\usepackage{dcolumn}
\usepackage{color}
\usepackage{xcolor}
\usepackage{float}
\usepackage{amsmath}
\usepackage{soul}
\usepackage{multirow}
\usepackage{booktabs}
\usepackage{url}
\usepackage{comment}

\begin{document}

\title{Low lying excitations in $^{150}$Pm}

\author{A.~Pal}
\affiliation{Variable Energy Cyclotron Centre, Kolkata - 700 064, India}
\affiliation{Homi Bhabha National Institute, Training School Complex, Anushakti Nagar, Mumbai - 400 094, India}

\author{S.~Basak}
\thanks{Present address: Institute of Modern Physics, Chinese Academy of Sciences, Lanzhou, Gansu - 730 000, China}
\affiliation{Variable Energy Cyclotron Centre, Kolkata - 700 064, India}
\affiliation{Homi Bhabha National Institute, Training School Complex, Anushakti Nagar, Mumbai - 400 094, India}

\author{T.~Bhattacharjee}
\thanks{Corresponding author}
\email{btumpa@vecc.gov.in}
\affiliation{Variable Energy Cyclotron Centre, Kolkata - 700 064, India}
\affiliation{Homi Bhabha National Institute, Training School Complex, Anushakti Nagar, Mumbai - 400 094, India}
\author{Nazira Nazir}
\affiliation{Department of Physics, University of Kashmir, Srinagar, 190006, India}
\author{G.~H.~Bhat}
\affiliation{Department of Higher Education (GDC Shopian), Shopian, Jammu and Kashmir 192303, India}
\author{S.~Jehangir}
\affiliation{Department of Higher Education (GDC Kulgam), Kulgam, Jammu and Kashmir 192231, India}
\author{D.~Kumar}
\affiliation{Variable Energy Cyclotron Centre, Kolkata - 700 064, India}
\affiliation{Homi Bhabha National Institute, Training School Complex, Anushakti Nagar, Mumbai - 400 094, India}
\author{A.~Saha}
\affiliation{Department of Physics, ICFAI University Tripura, Kamalghat, Tripura 799 210, India}
\author{S.~S.~Alam}
\affiliation{Government General Degree College, Chapra, West Bengal - 741 123, India}
\author{D.~Banerjee}
\affiliation{Variable Energy Cyclotron Centre, Kolkata - 700 064, India}
\affiliation{Homi Bhabha National Institute, Training School Complex, Anushakti Nagar, Mumbai - 400 094, India}
\author{A.~Das}
\affiliation{Variable Energy Cyclotron Centre, Kolkata - 700 064, India}
\author{A.~Adhikari} 
\affiliation{Ghani Khan Choudhury Institute of Engineering $\&$Technology, Malda - 732141, India}
\author{A.~Gupta}
\affiliation{Institute of Engineering \& Management, University of Engineering and Management, Kolkata - 700 091, India}
\author{S.~Das}
\affiliation{Institute of Engineering \& Management, University of Engineering and Management, Kolkata - 700 091, India}
\author{A.~Bisoi}
\affiliation{Indian Institute of Engineering Science and Technology, Shibpur, West Bengal - 711 103, India}
\author{Y.~Sapkota}
\affiliation{Dudhnoi College, Goalpara, Assam - 783 124, India}
\author{S.~Sharma}
\affiliation{Manipal University Jaipur, Jaipur-303007, Rajasthan, India}
\author{S.~Samanta}
\affiliation{Adamas University, Kolkata-700126, India }
\author{S.~Chatterjee}
\affiliation{UGC-DAE Consortium for Scientific Research, Kolkata-700 098, India}
\author{R.~Raut}
\affiliation{UGC-DAE Consortium for Scientific Research, Kolkata-700 098, India}
\author{S.~S.~Ghugre}
\affiliation{UGC-DAE Consortium for Scientific Research, Kolkata-700 098, India}
\author{J.~A.~Sheikh}
\affiliation{Department of Physics, University of Kashmir, Srinagar, 190006, India}

\date{\today}
\begin{abstract}
The low lying excitations in odd-odd $^{150}$Pm have been studied through proton induced reaction with an array of five Compton suppressed Clover HPGe and one segmented planar Ge detectors. The relative excitation functions for the observed $\gamma$ rays have been studied using singles data at two beam energies of 8~MeV and 9~MeV. 16 new $\gamma$ rays and 15 new levels have been placed in the level scheme of $^{150}$Pm based on $\gamma-\gamma$ coincidence data. The relative intensities for the observed $\gamma$ rays have been determined using total and gated projections. Tentative spin-parity assignments were made to few low lying excitations of $^{150}$Pm, using limited angular distribution data and other information. Lifetimes were estimated for two excited levels in this nucleus using using generalized centroid difference analysis, applied in the nanosecond range, with Ge detectors. Large basis shell model and projected shell model calculation were performed to interpret the experimentally observed levels. The present work indicates 1$^-$ ground state, a 2$^-$ state close to the ground state ($\sim$50~keV) and a low lying 6$^-$ isomeric state in this odd-odd nucleus along with emerging band structures developed with two quasiparticle configurations. 
\end{abstract}
\maketitle

\section{Introduction}
\label{intro}
Nuclear structure studies in the A $\sim$ 150 transitional nuclei, lying on the edge of the beta stability zone, are challenging yet extremely important. Even with advancements in state of the art experimental facilities and data analysis techniques, this domain remains very poorly studied. The scarcity of experimental data in this region is amply illustrated in the structural information of unstable Pm (Z = 61) isotopes~\cite{nndc}. 

The N = 89 $^{150}$Pm lies at the centre point of quantum shape phase transition (QPT) from a spherical vibrator to a deformed rotor at N = 88 to N = 90~\cite{iachello,casten}. Both the even-even N = 90 neighbors of $^{150}$Pm, viz. $^{150}$Nd~\cite{krucken} and $^{152}$Sm are known to exhibit X(5) symmetry features. The odd-A Pm were studied with interacting boson-fermion model (IBFM-2) model~\cite{barea} that showed signatures of transitions of vibrational to rotational core of the neighboring even-A isotopes. Therefore, level structure of $^{150}$Pm is important to understand the effect of shape transition and critical point symmetry in the odd - odd nuclei around N = 90.   

The existence of low lying isomers are expected in this nucleus due to coupling of the odd-proton and odd-neutron orbitals~\cite{vera} and this is also understood from the systematics~\cite{pmbeta}. The existence of isomers in these nuclei may significantly alter the nucleosynthesis through slow neutron capture process~\cite{pm148} and their identification is, therefore, of great importance. Identification from $\beta$ decay end point energy was attempted but was indecisive about the existence of any isomeric state in this nucleus~\cite{pmbeta}.

The excited states of $^{150}$Pm are of particular interest as they serve as intermediate states in the double $\beta$ decay of $^{150}$Nd to $^{150}$Sm, a process whose observation could shed some light on the Majorana nature of neutrinos~\cite{gauss}.

The sizeable population of the excited states of odd-odd $^{150}$Pm is challenging as only scattering reaction with light ions or fusion with proton and deuteron are the few routes that can be used with stable targets. As a result, populating higher angular momentum states is a difficult task. In addition, the spectroscopic measurements on this nucleus are limited with the yields of the de-excited $\gamma$ rays due to insufficient production cross section. The reaction cross section for proton induced fusion on $^{150}$Nd has been measured and the highest value was found to be $\sim$60~mb at 8~MeV~\cite{pmcs}. There is no $\beta$ decay route that populates $^{150}$Pm except through the double beta decay of $^{150}$Nd. The (n, $\gamma$) type of reaction are of not much use as Pm has no stable isotope. Owing to the above reasons, spectroscopic information about the excited states in the odd-odd nucleus $^{150}$Pm is very scanty. 

The Evaluated Nuclear Structure Data File (ENSDF)~\cite{ensdf} provides information on the ground state of $^{150}$Pm , assigning a possible spin–parity of (1$^{-}$) and a half-life of 2.698(15)~h, while excited states up to 711.8~keV are reported in the database. In one of the recent works~\cite{gauss}, the $^{150}$Nd($^3$He,t) reaction was used to populate the states of $^{150}$Pm and a 2$^-$ ground state was proposed with an assumption on the absence of any close lying 1$^-$ state near the ground state. Therefore, it is of primary importance to understand the correct spin-parity of the ground state for this nucleus. 

The rudimentary level structure of $^{150}$Pm, proposed to be developed on the ground state, is available in literature~\cite{bucu} from charge exchange type of experiments. In this work, the levels have been studied using (p,n$\gamma$) and (d,$\alpha$) reactions and compared with the ($^3$He, t) data~\cite{gauss}. The $\gamma-\gamma$ coincidence information were mainly derived from the data gathered with four planar high purity Ge detectors in presence of a 1liter NE215 scintillator detector for neutrons. 

In the present work, the low lying level structure of odd-odd $^{150}$Pm has been studied using an array of five BGO suppressed Clover HPGe detectors and a segmented planar Ge detector ( also known as Low Energy Photon Spectrometer (LEPS)), setup at the Channel III beam line of K-130 cyclotron at VECC, Kolkata. The large basis spherical shell model calculation has been performed to understand the low lying levels and projected shell model calculation has been utilized to interpret the experimental observation in terms of quasiparticle excitations. 

\section{Experiment}
\label{expt}
The low lying excited states of $^{150}$Pm were populated through $^{150}$Nd(p,n)$^{150}$Pm reaction with proton beam delivered from K-130 cyclotron at VECC, Kolkata. 97.65\% enriched $^{150}$Nd target was prepared using centrifuge technique with a thickness $\sim$10mg/cm$^2$. Data were gathered at two different beam energies of 8 MeV and 9 MeV. The decaying $\gamma$-rays from the excited states of the residues were detected using Indian National Gamma Array setup, comprising of five Compton suppressed clovers (two at 125$^{\circ}$(2) and three at 90$^{\circ}$) and one LEPS at 40$^{\circ}$ at the time of the experiment. The data were recorded in list mode using PIXIE-16 digitizers. The list mode data satisfying one Clover hit was written and were also sorted using the program IUCPIX~\cite{sdas} to generate $\gamma-\gamma$ matrix. The $\gamma$ singles and $\gamma-\gamma$ coincidence data were subsequently analyzed using the program INGASORT~\cite{ingasort}. 

\section{Data Analysis and Results}
This section explains the development of low lying level structure of $^{150}$Pm that has been performed in the present work. The $\gamma$ rays were assigned in the level scheme using relative excitation function and $\gamma -\gamma$ coincidence information. The symmetric (Clover-Clover) and assymetric (Clover-LEPS) 4K $\times$ 4K matrices, developed with a 200~ns coincidence time window, were used to extract the coincidence relations. The spin assignments could be made for few levels using angular distribution analysis and other available information. The relative intensities of the transitions were found for the first time in the present work using the total and gated projection of the $\gamma-\gamma$ matrix. In addition, limits on lifetime of two levels have been estimated using centroid shift method.

\begin{figure*}[ht!]
\begin{center}
\includegraphics[width=\textwidth, height=7cm]{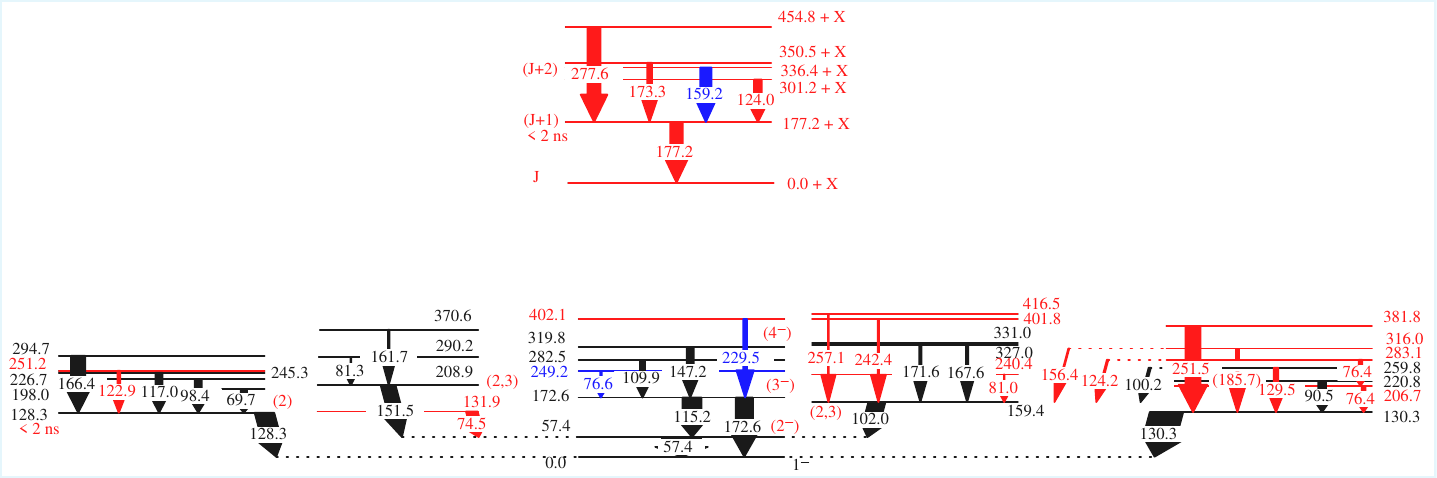} 
\caption{(Color online) Low lying level structure of $^{150}$Pm proposed in the present work. The newly found $\gamma$ rays, levels and spin-parities and halflives (T$_{1/2}$) are shown in red. The $\gamma$ rays or levels for which the position is altered compared to earlier work~\cite{bucu} are shown in blue. The relative intensities have been measured in the present  work and indicated in the figure with the width of the transitions. No conversion has been considered for the transitions, while drawing the level scheme, as their multipolarities are not confirmed.}
\label{level}
\end{center}
\end{figure*}

In the work of Ref.~\cite{bucu}, the $\gamma$ rays were assigned to the levels scheme of $^{150}$Pm from the (p,n) reaction at 7.1~MeV, considering that no $^{149}$Pm will be produced. This assumption corroborates with the observation of our earlier work~\cite{pmcs}. However, very small yield of $^{149}$Pm was indicated in the work of Lebeda et al.~\cite{lebeda} with a comparatively smaller error bar in their data. Also, from both these works, it has been found that the yield of $^{150}$Pm from (p,n) reaction is highest at 8.0~MeV. Therefore, all the coincidence and intensity analyses were done using the data with 8~MeV proton beam. The data at 9~MeV was used only for excitation function analysis. The following sections describe the details on these analysis procedures and results. The low lying level scheme of $^{150}$Pm, proposed in the present work, has been shown in Fig.~\ref{level} where the newly assigned levels, $\gamma$ rays and spin-parities are shown with red. 

\subsection{Relative excitation function and $\gamma$ rays in $^{150}$Pm}
\label{rg}

Cross section data~\cite{pmcs,lebeda} shows that the yield of $^{150}$Pm and $^{149}$Pm is almost similar at 8~MeV and the yield of $^{149}$Pm is more than three times larger at 9~MeV compared to that at 8~MeV. In addition, it is also seen that no other residue is produced at these beam energies except $^{149,150}$Pm. These information could be used for assignment of $\gamma$ rays to $^{150}$Pm comparing the singles $\gamma$ yields at 8~MeV and 9~MeV. 

To quantify the change in $\gamma$-ray yields between the two beam energies (8 and 9~MeV), a normalization factor was introduced based on the relative yields of a known transition in $^{149}$Pm, specifically the 211~keV line. This factor $F = \frac{Y_{211}^{8MeV}}{Y_{211}^{9MeV}} = \frac{\phi _{8}\sigma _8 t_8 ^{irr}}{\phi _{9}\sigma _9 t_9 ^{irr}}$ accounts for variations in experimental conditions, such as beam intensity ($\phi$) and irradiation time ($t^{irr}$), between the two measurements, in addition to the production cross sections ($\sigma$). 

Applying this normalization allows for a consistent comparison of the $\gamma$-ray yields of other transitions observed in the spectra. Based on this methodology, normalized yield ratios were determined for all observed transitions, following the relations given below.
\begin{eqnarray}
\nonumber
R_{\gamma}& = &\frac{Y_{\gamma}^{8MeV}}{F \times Y_{\gamma}^{9MeV}}\\ \nonumber
&=&\frac{Y_{211}^{9MeV}\times Y_{\gamma}^{8MeV}}{Y_{211}^{8MeV}\times Y_{\gamma}^{9MeV}}\\ \nonumber
&=&\frac{(\phi _{9MeV}t^{irr}_{9MeV}\sigma^{^{149}Pm} _{9MeV})\times(\phi _{8MeV}t^{irr}_{8MeV}\sigma _{8MeV})}{(\phi _{8MeV}t^{irr}_{8MeV}\sigma^{^{149}Pm} _{8MeV})\times(\phi _{9MeV}t^{irr}_{9MeV}\sigma _{9MeV})}\\
&=&(\frac{\sigma^{^{149}Pm}_{9MeV} }{\sigma _{9MeV}}) \times (\frac{\sigma _{8MeV}}{\sigma^{^{149}Pm}_{8MeV}})
\end{eqnarray} 
where, $Y_{\gamma}$ is the area of the observed transitions in the singles projection of two different beam energies, $\phi$, $t^{irr}$ and $\sigma$ are same as described before. It may be assumed that all transitions originating from $^{149}$Pm will exhibit similar normalized yield ratios close to unity, since their production cross sections and experimental conditions do not vary significantly. On the other hand, transitions associated with $^{150}$Pm, which is produced more abundantly at 8~MeV, are expected to show noticeably higher normalized ratios after correction.

It was found that the transitions unambiguously assigned to $^{149}$Pm indeed showed ratios close to 1.0, validating the approach. In contrast, transitions belonging exclusively to $^{150}$Pm exhibited much higher yield ratios, generally greater than 3.3, consistent with the expected cross-section behavior for the (p,n) reaction at these beam energies.

\subsection{$\gamma -\gamma$ coincidence}
\label{ls}
\begin{figure}[ht!]
\begin{center}
\includegraphics[width=\columnwidth]{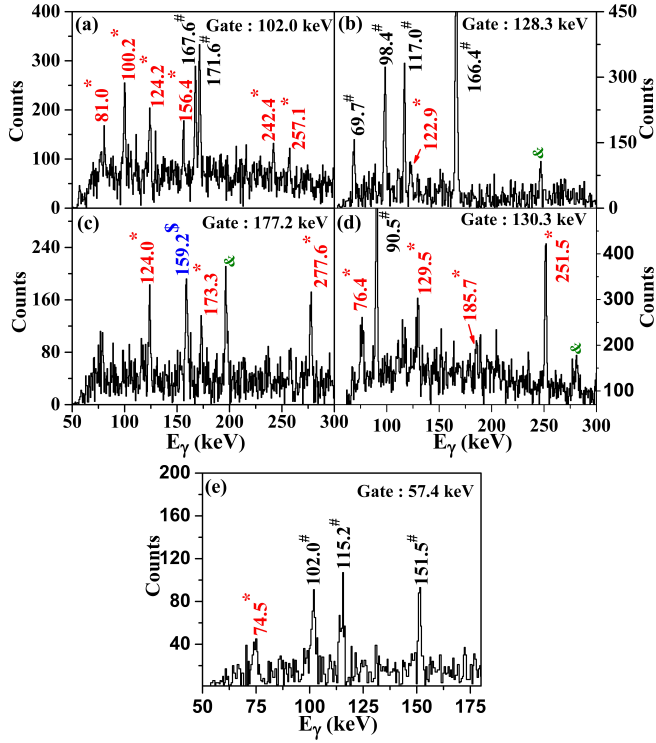}
\caption{(Color online) The $\gamma-\gamma$ coincidence spectra from Clover-Clover (a-d) and LEPS-Clover (e) matrices are shown for few selected transitions, indicating the placement of new $\gamma$ rays (indicated with red $\ast$ ) in the level scheme. The $\gamma$ rays, indicated with blue and marked with \$ have altered position compared to earlier work and the one indicated in green and marked with \& could not be placed in the level scheme. The $\gamma$ transitions known earlier and confirmed in the present work are marked with \#.}
\label{gates}
\end{center}
\end{figure}

 The level scheme of $^{150}$Pm  presented in Fig.~\ref{level} has been constructed after a detailed and careful $\gamma\gamma$ coincidence analysis, as several $\gamma$ rays with closely spaced energies, belonging to both $^{149}$Pm and $^{150}$Pm, were observed in the spectrum. The coincidence relationships of the observed $\gamma$ rays which showed R$_{\gamma}$ substantially greater than 1.0 were studied from the 4K $\times$ 4K Clover-Clover matrix generated from the data gathered with 8~MeV proton. The LEPS vs Clover matrix was also used to check the coincidences of the low energy $\gamma$ lines. Few representative gates are shown in  Fig.~\ref{gates}, displaying the coincidence relationships.
The new $\gamma$ lines placed in the level scheme are indicated in red and marked with $\star$ in Fig.~\ref{gates}. All the newly found transitions are also shown in red in Fig.~\ref{level}. Three $\gamma$ rays, viz., 76.6~keV, 159.2~keV and 229.5~keV, were found to have different position compared to the earlier placements and are marked in blue in Fig.~\ref{level}. The (185.7)~keV $\gamma$ line has been tentatively placed in the level scheme as this transition is clearly seen in the 130.3~keV gate but the cross gate on 185.7~keV could not be examined as this transition is not clearly seen in the total projection. Similarly, a very weak indication of one 258.7~keV transition is seen in the 177.2~keV gate but was not placed in the level scheme. The list of $\gamma$ rays that belong to the level scheme of $^{150}$Pm are also listed in Table~\ref{tab:combined_gamma_with_intensity_shifted} along with the initial and final level energies for their decay. 
\begin{table}[ht!]
\centering
\scriptsize
\renewcommand{\arraystretch}{1}
\caption{Details on the placement of the $\gamma$ rays in the level scheme of $^{150}$Pm, their intensities, angular distribution coefficients (when available) and spin parity assignments (conjectured either from angular distribution data or otherwise) are shown. 
The intensities are determined with an error of less than 5\% for strong transitions and it was about 25\% for the weakest ones.}
\begin{tabular}{|c|c|c|c|c|}
\hline
\textbf{E$_\gamma$ (keV)} & \textbf{I$_\gamma$} & \textbf{a$_2$} & \textbf{E$_i \rightarrow$ E$_f$ (keV)} & \textbf{J$_i^{\pi} \rightarrow$ J$_f^{\pi}$} \\
\hline
57.4  & 37.90 & - & $57.4 \rightarrow 0.0$ & $(2^-) \rightarrow 1^-$ \\
69.7  & 6.01 & - & $198.0 \rightarrow 128.3$ & $ \rightarrow (2)$ \\
74.5  & 11.01 & - & $131.9 \rightarrow 57.4$ &  $ \rightarrow (2^-)$  \\
76.4  & 3.88 & - & $206.7 \rightarrow 130.3$ & - \\
76.4  & 3.88 & - & $283.1 \rightarrow 206.7$ & - \\
76.6  & 1.49 & - & $249.2 \rightarrow 172.6$ & $ \rightarrow (3^-)$  \\
81.0  & 0.56 & - & $240.4 \rightarrow 159.4$ & $ \rightarrow (2,3)$ \\
81.3  & 0.75 & - & $290.2 \rightarrow 208.9$ & $ \rightarrow (2,3)$  \\
90.5  & 7.62 & - & $220.8 \rightarrow 130.3$ & - \\
98.4  & 6.68 & - & $226.7 \rightarrow 128.3$ & $ \rightarrow (2)$  \\
100.2 & 2.16 & - & $259.6 \rightarrow 159.4$ & $ \rightarrow (2,3)$  \\
102.0 & 16.05 & -0.26(3) & $159.4 \rightarrow 57.4$ & $(2,3) \rightarrow (2^-)$ \\
109.9 & 5.09 & - & $282.5 \rightarrow 172.6$ & $ \rightarrow (3^-)$  \\
115.2 & 18.57 & - & $172.6 \rightarrow 57.4$ & $(3^-) \rightarrow (2^-)$ \\
117.0 & 6.77 & - & $245.3 \rightarrow 128.3$ & $ \rightarrow (2)$  \\
122.9 & 2.61 & - & $251.2 \rightarrow 128.3$ & $ \rightarrow (2)$  \\
124.0 & 7.68 & - & $301.2+X \rightarrow 177.2+X$ & $ \rightarrow (J+1)$  \\
124.2 & 1.41 & - & $283.1 \rightarrow 159.4$ & $ \rightarrow (2,3)$  \\
128.3 & 18.44 & -0.24(6) & $128.3 \rightarrow 0.0$ & $(2) \rightarrow 1^-$ \\
129.5 & 4.50 & - & $259.8 \rightarrow 130.3$ & - \\
130.3 & 31.42 & - & $130.3 \rightarrow 0.0$ & $ \rightarrow 1^-$ \\
147.2 & 6.90 & -0.85(5) & $319.8 \rightarrow 172.6$ & $(4^-) \rightarrow (3^-)$ \\
151.5 & 14.03 & -0.63(9) & $208.9 \rightarrow 57.4$ & $(2,3) \rightarrow (2^-)$ \\
156.4 & 0.92 & - & $316.0 \rightarrow 159.4$ & $ \rightarrow (2,3)$  \\
159.2 & 10.99 & -0.76(10) & $336.4+X \rightarrow 177.2+X$ & $(J+2) \rightarrow (J+1)$ \\
161.7 & 1.46 & - & $370.6 \rightarrow 208.9$ & $ \rightarrow (2,3)$  \\
166.4 & 13.85 & - & $294.7 \rightarrow 128.3$ & $ \rightarrow (2)$  \\
167.6 & 2.16 & - & $327.0 \rightarrow 159.4$ & $ \rightarrow (2,3)$  \\
171.6 & 2.33 & - & $331.0 \rightarrow 159.4$ & $ \rightarrow (2,3)$  \\
172.6 & 12.22 & - & $172.6 \rightarrow 0.0$ & $(3^-) \rightarrow 1^-$ \\
173.3 & 4.53 & - & $350.5+X \rightarrow 177.2+X$ & $ \rightarrow (J+1)$  \\
177.2 & 12.04 & -0.60(8) & $177.2+X \rightarrow 0.0+X$ & $(J+1) \rightarrow J$ \\
(185.7) & 3.37 &  & $316.0 \rightarrow 130.3$ & - \\
229.5 & 3.68 & - & $402.1 \rightarrow 172.6$ & $ \rightarrow (3^-)$  \\
242.4 & 1.43 & - & $401.8 \rightarrow 159.4$ & $ \rightarrow (2,3)$  \\
251.5 & 14.55 & - & $381.8 \rightarrow 130.3$ & - \\
257.1 & 0.70 & - & $416.5 \rightarrow 159.4$ & $ \rightarrow (2,3)$  \\
277.6 & 12.68 & - & $454.8+X \rightarrow 177.2+X$ & $ \rightarrow (J+1)$  \\
\hline
\end{tabular}
\label{tab:combined_gamma_with_intensity_shifted}
\end{table}
\begin{table}[ht!]
\centering
\scriptsize
\renewcommand{\arraystretch}{1.2}
\caption{Comparison of the experimental level energies determined in the present work are compared to earlier known experimental data from $(p,n\gamma)$, $(d, \alpha)$, and $(^3\text{He},t)$ reactions.}
\begin{tabular}{|c|c|c|c|}
\hline
\textbf{Pres. Exp. (keV)} & \textbf{( p,n$\gamma$ ) ~\cite{bucu}} & \textbf{( d,$\alpha$ ) ~\cite{bucu}} & \textbf{( $^3$He,t ) ~\cite{gauss}} \\
\hline
0.0     &  0.0       & 0.0      & 0.0 \\
        &          &             &     \\
57.4    &  56.8    & 57.1 (7)    &     \\
        &   115.8   & 112.4 (11)  & 110 \\
128.3   &    127.9   & 125.7 (10)  &  \\
130.3   &     129.7      &              &     \\
131.9   &        &            &     \\
159.4   &     158.6   & 157.4 (13)  &     \\
172.6   &  172.3   & 171.5 (8)   &     \\
177.2+X &          &            &     \\
198.0   &  196.5   &            & 190 \\
206.7   &          &             &     \\
208.9   &  208.0    & 207.9 (15)  &     \\
220.8   &  219.9   &            &     \\
226.7   &  225.9   &            &     \\
        &  229.5   & 228.3 (15)  &     \\
240.4   &          &             &     \\
245.3   &   244.4   &            &     \\
249.2   &  (247.2) & 249.5 (10)  &     \\
251.2   &          &            &     \\
259.8   &  258.5  &       &  \\
   &    & 264.4 (6)      &  \\
282.5   &  281.7   &            & 280 \\
283.1   &           &            &     \\
290.2   &   288.8  & 287.6 (10) &     \\
294.7   &   293.9  &           &     \\
301.2+X &          &            &     \\
        &          & 304.1 (6)  &     \\
316.0   &          &            &     \\
319.8   &  319.2    & 320.8 (7)  &     \\
327.0   &   326.0  &           &     \\
331.0   &   330.0    & 331.8 (11) & \\
336.4+X &          &            &   \\
350.5+X &           &           &  \\
        &           &  366.5 (10)&  \\
370.6   &   369.3  &           &     \\
381.8   &         &           &     \\
        &         & 388.9 (22) &     \\
401.8   &         & 401.2 (12) & 400  \\
402.1   &          &        &     \\
416.5   &         & 417.2 (6) &     \\
        &          & 431.7 (15) &     \\
        &          & 442.3 (8)  &     \\
454.8+X &          &             &     \\
        &    & 471.6 (7)   &     \\
\hline
\end{tabular}
\label{tab:energy_compare}
\end{table}

Two sets of $\gamma$ rays were found in the present work that could be assigned to the level scheme of $^{150}$Pm. One set can be placed on the ground state based on the existing information and the coincidence relations connecting all these $\gamma$ lines. These $\gamma$ rays form one part of the level scheme of $^{150}$Pm and are shown in the bottom panel of Fig.~\ref{level}.

Among the second set of $\gamma$ lines, which have no connection with $\gamma$ rays of the first set, the 128.3~keV transition and the $\gamma$ rays in coincidence with 128.3~keV were placed on the ground state, considering the earlier propositions by Bucurescu et al., based on both (p, n$\gamma$) and (d,$\alpha$) reactions. However, the 177.2~keV and the transitions which are in coincidence with 177.2~keV, are tentatively placed on the 6$^-$ isomer in $^{150}$Pm (shown on top of Fig.~\ref{level}), based on the shell model prediction shown in section~\ref{sm}. However, no confirmation on such assumption could be made in the present work.

The presence of 112.4~keV level~\cite{bucu} could not be confirmed in the present work as the 57.4~keV gate in LEPS detector does not show another $\sim$57~keV transitions in the Clover projection from LEPS-Clover matrix. This could be due to high energy threshold greater 50~keV set in the Clover detectors that was maintained in the campaign. The level energies proposed from the present work are shown in Table~\ref{tab:energy_compare} in comparison to the earlier information known in literature~\cite{gauss,bucu}.

\subsection{Intensity measurements}
\label{intensity}

The relative intensities of $\gamma$ rays in $^{150}$Pm have been estimated for the first time in the present work. Due to the presence of many closely spaced $\gamma$ rays and contributions from $^{149}$Pm, both total projection and gated energy projections were used to ensure accurate intensity determination of these $\gamma$ rays. The Table~\ref{tab:combined_gamma_with_intensity_shifted} also lists the intensities of $\gamma$ rays placed in the level scheme. During this calculation, it has been considered that sum of the intensities of all the transitions feeding to the ground state adds up to 100. 

 Intensity mismatch were found for the 128.3~keV and 177.2+X~keV levels de-exciting by the 128.3~keV and 177.2~keV transitions. This can be explained either with an appropriate conversion coefficients or with delayed decay for these two levels. The estimation of conversion coefficient for these two transitions will require the difference between absolute feeding (I$_{feeding}$) and decaying (I$_{decay}$) intensities. As most of the feeding transitions are low energy $\gamma$ rays, it would be important to know their conversion coefficients, in addition to knowing any other unobserved feeding or decaying transitions, to estimate the accurate conversion for the decaying transitions. Even ignoring the conversion of the feeding transitions and/or the missing transitions, only the $\gamma$ intensity mismatch at these levels suggests conversion coefficients ($\alpha = \frac{I_{feeding}}{I_{decay}} - 1$) of 1.0 and 1.97, respectively, for the 128.3~keV and 177.2~keV transitions. 
 
  This can be realized if the 128.3~keV $\gamma$ ray is
 E1+M2 ($\delta$ = 0.4, $\alpha _{BRICC}$ = 0.96) in nature, as understood from the conversion coefficients, calculated using BRICC code~\cite{bricc}. This conjecture also fits with the angular distribution and lifetime analysis, shown below, made for this transition. In case of 177.2~keV transition, the multipolarity E1+M2 ($\delta$ = 5,  $\alpha _{BRICC}$ = 1.83), E2+M3 ($\delta$ = 0.45,  $\alpha _{BRICC}$ = 1.97) or stretched E3 ($\alpha _{BRICC}$ = 1.99) may justify the observed intensity mismatch. The E1+M2 indicates the level lifetime in the order of picoseconds for both the levels whereas the later two multipolarities will lead to a lifetime for the 177.2+X~keV level to be $\sim$$\mu$s as per Weisskopf estimate. However, as reflected in the angular distribution and lifetime analysis presented below the E1+M2 assignment seems to be more feasible for both the transitions.   
 
 \subsection{Spin-Parity (J$^{\pi}$) assignment}
\label{sp}

Attempt has been made for tentative J$^{\pi}$ assignment using angular distribution analysis wherever possible and few other logical parameters towards conjecturing the J$^{\pi}$ values. For few cases, conjectures could be made from comparison with shell model calculation. Following sections describe the spin-parity assignment to ground state, first excited state and few other excited levels in the nucleus.

\subsubsection{Ground state}
\label{gs}

The spin-parity of 1$^-$ was conjectured, for the ground state of $^{150}$Pm, by Hoshi et al.,~\cite{hoshi} from the measurement of {\it logft} values. This was determined through the measurement of intensity of $\beta$ decay of $^{150g}$Pm to $^{150g}$Sm. This is the only measurement of this particular ground state to ground state beta decay intensity which resulted in the assignment of ground state J$^{\pi}$ of $^{150}$Pm as 1$^-$. This follows the proposition made in the work of Barret et al.,~\cite{barret} for the intensity for the $^{150g}$Pm $\rightarrow$ $^{150g}$Sm $\beta$ decay as $<$3\% and ground state J$^{\pi}$ of $^{150}$Pm as 1$^-$. 

Recently, the $^{150}$Nd($^3$He,t) reaction was used to populate $^{150}$Pm~\cite{gauss} and a 2$^-$ ground state for this nucleus  was proposed with an assumption on the absence of any close lying 1$^-$ state near the ground state in $^{150}$Pm. So, it is possible that the 2$^-$ state has been excited through the $^{150}$Nd($^3$He,t) reaction and might have been interpreted as the ground state due to very small energy difference of this level from the 1$^-$ ground state that was not observed owing to the high energy resolution of the charged particle detector of about 40~keV. The presence of a close lying 2$^-$ state (31~keV) above the 1$^-$ ground state is indeed predicted in the shell model calculation, as discussed in section~\ref{sm}.

The 2$^-$ assignment of the ground state was adopted by Bucurescu et al.~\cite{bucu} from the work of Gauss et al~\cite{gauss} and this was further validated from the {\it logft} measurements~\cite{pm_arunabha}. This later work relied on the evaluated data on $^{150}$Pm~\cite{ensdf} and considered a $\beta$ decay intensity of 10\% for $^{150g}$Pm $\rightarrow$ $^{150g}$Sm decay, in contrast to the measured value of 1.04\% ~\cite{hoshi} for entire {\it logft} calculation. Therefore, this can be ignored for the present discussion. 

Having all the above observations in hand, the following conclusion may be outlined for the ground state in $^{150}$Pm. Firstly, the {\it logft} value determined from the measured intensity for the ground state to ground state $\beta$ decay~\cite{hoshi} may be considered as the most reasonable experimental data to conjecture the J$^{pi}$ = 1$^-$ for the ground state of $^{150}$Pm which is also validated from {\it logft} measurements~\cite{barret}. Secondly, the existence of a extremely close lying 2$^-$ state can not be ignored following the observations in ($^3$He,t) data~\cite{gauss}. One can depend on the predictions of the shell model shown in Section~\ref{sm} with 1$^-$ as ground state and 2$^-$ as the close lying excited state, as the calculation could accurately determine the known ground state J$^{\pi}$ values of the neighboring odd-Z ($^{149}$Pm), odd-N ($^{149}$Nd) and odd-odd ($^{148}$Pm) nuclei. 

Therefore, in the present work, the ground state of $^{150}$Pm is assigned to have the spin-parity of 1$^-$.

\subsubsection{57.4~keV excited state}
\label{57}

This 57.4~keV state is found to be the first excited state in $^{150}$Pm. However, the determination of multipolarity for the 57.4~keV transition has not been possible from any conventional method. In the present work, J$^{\pi}$ of this level is conjectured to be 2$^-$ following the prediction of shell model which has successfully reproduced the first excited state in neighboring $^{148}$Pm (section~\ref{sm}). In addition, it is observed from the shell model calculation that the first positive parity level is at much higher excitation (above 1~MeV) compared to the negative parity levels. Accordingly, positive parity assignment for this level was ignored. 

\subsubsection{Other excited levels}

\begin{figure}[ht!]
\begin{center}
\includegraphics[width=\columnwidth]{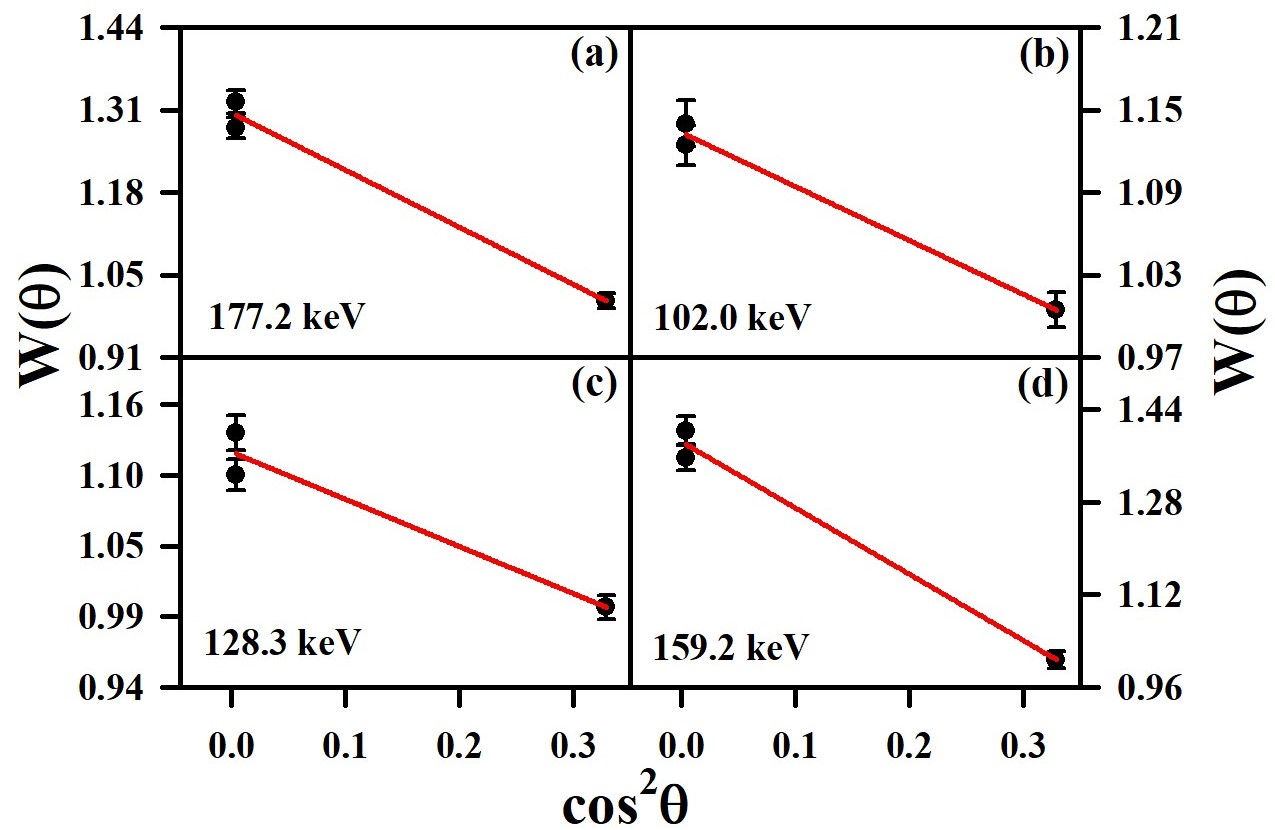}
\caption{(Color online) The results from angular distribution analysis are shown for few transitions in $^{150}$Pm. The black circles represent the experimental data points and the red lines indicate the linear fit.}
\label{fig_ang}
\end{center}
\end{figure}

The spin assignment for few excited levels of $^{150}$Pm were attempted from a limited angular distribution analysis. In order to realize this measurement, the 90$^{\circ}$ detectors were divided into two halves, thereby, generating data at two different angles of 86.5$^{\circ}$ and 93.5$^{\circ}$. This could be done due to the known opening angle of the Clover HPGe detectors as 7$^{\circ}$ following the correctness of the procedure already demonstrated~\cite{sm152}. With data in three different angles (86.5$^{\circ}$, 93.5$^{\circ}$ and 125$^{\circ}$), the angular distribution analysis was performed by plotting W($\theta$) as a function of cos$^2 (\theta)$ which provided the angular distribution coefficient a$_2$. Although, both a$_2$ and a$_4$ coefficients are required to determine the correct multipolarity of a transition, the sign of the a$_2$ value can guide about the gross nature (dipole or quadrupole)~\cite{sm152}. The experimental a$_2$ values, determined from angular distribution analysis, are represented in Fig.~\ref{fig_ang}. This shows that most of the transitions have W($\theta$) downsloping with increasing $\theta$ giving rise to $a_2 < 0$. This indicated that the analysed $\gamma$ rays are either dipole or mixed but not stretched E2s or E3s (for which $a_2 > 0$) in nature~\cite{zolt}.
 
\begin{figure}[ht!]
\begin{center}
\includegraphics[width=\columnwidth]{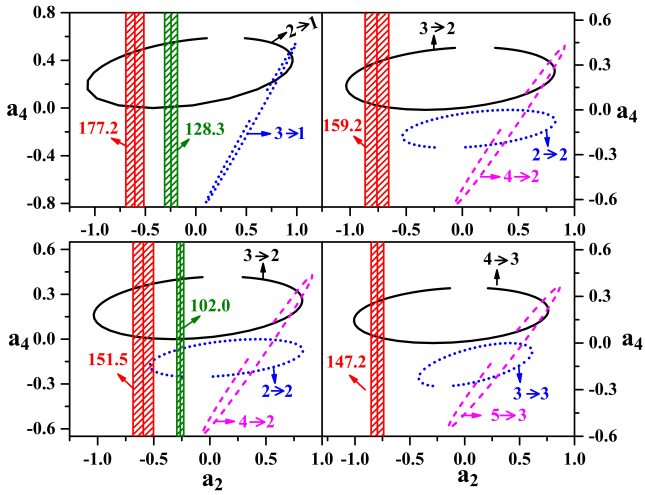}
\caption{(Color online) The contours have been drawn from the a$_2$ and a$_4$ values obtained from ANCORR calculations. The experimental a$_2$ values have been shown with verical bars, in comparison to these contours. The errors in a$_2$ values are indicated by the horizontal width of the shaded regions.}
\label{contour2}
\end{center}
\end{figure}

The experimental a$_2$ values have been compared with the a$_2$ vs a$_4$ contours calculated with ANCORR code~\cite{ancor}, as shown in Fig.~\ref{contour2}. To calculate the theoretical values of the angular distribution coefficients a$_2$ and a$_4$, one must consider degree of alignment of the excited state, which is characterized by the $\sigma /J$ ratio. It is found that $\sigma/J \approx 0.3$ reproduces the a$_2$ coefficients reported for pure dipoles~\cite{ang1} that used the same reaction used in the present work for population of $^{149}$Pm. This was also found that the same  $\sigma/J$ ratio was used for calculation of a$_2$ and a$_4$ contours in other experiments using (p,n) reaction~\cite{ang2}. This value of $\sigma /J$ has therefore been adopted in the present work for the theoretical calculations a$_2$ and a$_4$ coefficients using ANCORR.

During present analysis, only J $\rightarrow$ J$^<$(less than J) and in some cases, J $\rightarrow$ J transitions were considered by comparing the a$_2$ values with the a$_2$ vs a$_4$ contours, shown in Fig.~\ref{contour2}. These results, however, could indicate spin assignments for very few levels in $^{150}$Pm and are discussed below.

The $a_2$ value could not be determined for the 172.6~keV transition. It is found that the 57.4~keV and 115.2~keV transitions are placed as decaying from the 172.6~keV level in parallel to 172.6~keV transition. Considering the (2$^-$) nature for the 57.4~keV level, the 172.6~keV level was considered to have the assignment of (3$^-$) which explains the placement of 172.6~keV transition as a cross over to the 57.4~keV and 115.2~keV transitions. Figure~\ref{contour2} shows that the a$_2$ value for 147.2~keV $\gamma$ ray fits with the $4 \rightarrow 3$ assignment. More experimental data is necessary to make firm assignments on the spin parities of these levels.

The a$_2$ values determined for the 151.5~keV and 102.0~keV transitions, in comparison to the a$_2$ vs a$_4$ contours, show that the 208.9~keV level can have an assignment of J = 3 and a J = 2,3 assignment can be made for the 159.4~keV level. It is indeed not possible to conjecture parities for these levels. 

The 177.2~keV, 159.2~keV and 128.3~keV transitions show negative a$_2$ values. The 128.3~keV level can have J = 2 following the contour plot of Fig.~\ref{contour2}. This fits with the conjecture on the multipolarity of this transition from observed intensity mismatch. In this figure, the a$_2$ values of 159.2~keV and 177.2~keV transitions are also shown in comparison to the contours for dipoles, although no spin assignments were made for the levels decaying these transitions, considering that they might be placed on the 6$^-$ isomer in this nucleus. It can be seen that both the 159.2~keV and 177.2~keV $\gamma$ rays can be assigned with J $\rightarrow$ J-1 character. This conjecture also fits with the E1+M2 multipolarity of the 177.2~keV transition with a high M2 mixing, as would have been required to justify the observation of intensity mismatch for 177.2+X~keV level. 

The determined a$_2$ values along with the tentative spin assignments for the levels in $^{150}$Pm are given in Table~\ref{tab:combined_gamma_with_intensity_shifted} along with other relevant information. Future measurements on angular distribution from fusion evaporation or transfer reaction data would be extremely important for more definite spin parity assignments.

\subsection{Lifetime analysis}
\label{lt}

The lifetime information is important for the levels decaying the 128.3~keV and 177.2~keV transitions following the observation of considerable intensity mismatch. The lifetime measurements with the present experimental setup was attempted following the centroid shift ($\Delta$C) of the delayed time spectrum of a particular cascade (E$_{\gamma}(feeding)$ - E$_{\gamma}(decay)$) with respect to the anti-delayed time spectrum of the same cascade~\cite{devesh,shefali}. Similar measurements are most often performed with the fast timing scintillator detectors to determine nuclear level lifetime in the order of tens of picoseconds. However, the technique of centroid shift is, ideally, applicable to any $\gamma$ detection setup with reasonable energy resolution and also for lifetimes in the range of few nanoseconds to few tens of nanoseconds~\cite{ajay,sujit}. However, often in such cases, a prompt time calibration at the energy of interest remains unattended.

In Generalized Centroid Difference (GCD) technique~\cite{regis2016}, the time difference between the delayed and anti-delayed distribution is compared with the prompt response of the setup at the energy difference ($\Delta$E = E$_{\gamma 1}$ - E$_{\gamma 2}$) of interest, corresponding to the particular cascade. The mathematical expressions used in this measurement are given below for ready reference to the readers. 
\begin{figure}[ht!]
\begin{center}
\includegraphics[width=\columnwidth]{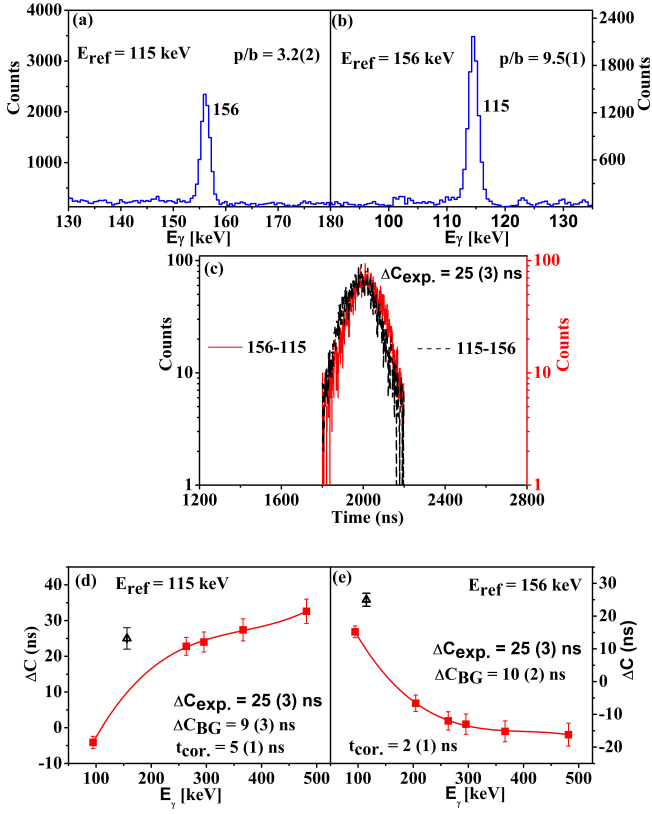}
\caption{(Color online) The timing analysis with the 156-115~keV cascade of $^{149}$Pm is shown with the gated projections (a,b); which also shows the {\it p/b} ratios, the delayed (red solid line) and anti-delayed (black dashed line) time distributions (c); this shows the centroid difference ($\Delta C_{expt}$ = 25(3) ns obtained from the analysis and the background correction plots (with red squares and solid lines) for feeder (d) and decay (e), respectively. The  $\Delta C_{BG} (feeder)$ = 9(3)~ns $\Delta C_{BG} (feeder)$ = 10(2)~ns were utilized to calculate the background correction in time t$_{corr}$ = 2(1)~ns, utilizing the {\it p/b} ratios. The corrected centroid difference $\Delta C_{FEP}$ = 29(3)~ns was, therefore, determined from the $\Delta C_{expt}$ (triangles in (d) nd (e) ). This, in turn, gives rise to a PRD value of 24(3)~ns at 156~keV, in comparison to PRD=0~ns at 115~keV, using the known halflife  value of 2.53(3)~ns .}
\label{fig_pr_decay1}
\end{center}
\end{figure}
\begin{figure}[ht!]
\begin{center}
\includegraphics[width=\columnwidth]{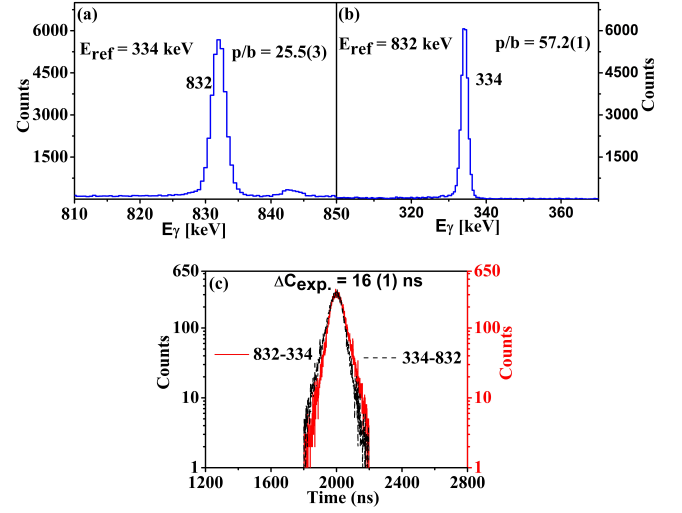}
\caption{(Color online) The timing analysis with the 832-334~keV cascade of $^{150}$Sm is shown with the gated projections (a,b); which also shows the {\it p/b} ratios, the delayed (red solid line) and anti-delayed (black dashed line) time distributions (c); this shows the centroid difference ($\Delta C_{expt}$ = 16(1) ns obtained from the analysis. This particular cascade does not require any background correction owing to very high {\it p/b} ratio greater than 20. Therefore, the corrected centroid difference $\Delta C_{FEP}$ remains same as = $\Delta C_{expt}$ = 16(1)~ns. The PRD value of 832~keV was determined to be 84(2)~ns with respect to PRD = 0 at 115~keV using the known halflife value of 0.05(3)~ns, following the procedure described in Ref.~\cite{regis2016}.}
\label{fig_pr_decay2}
\end{center}
\end{figure}
The time difference distribution detected with a particular cascade of two $\gamma$ rays is given by $N(t) = N_0 \lambda \int_{-\infty}^{t} P(t^{\prime}-t_0) e^{-\lambda (t^{\prime} - t_0)}dt^{\prime}$, which is a convolution of the prompt response function P(t) of the setup, measured for the same $\gamma$ energies of the cascade, with the exponential decay of the intermediate nuclear level with lifetime $\tau$. The centroid position of this distribution (C$_d$) depends on the centroid position of P(t) (C$pd$) and lifetime of the intermediate level ($\tau$); and can be expressed as $C_d = C_{pd} +\tau$. With the same cascade, one can also determine the anti-delayed time difference distribution for which $C_{ad} = C_{pad} - \tau$. As a result, the relative centroid difference for the delayed and anti-delayed time difference distributions are given by $\Delta C = C_d - C_{ad} = 2\tau + (C_{pd} - C_{pad})$. The quantity $(C_{pd} - C_{pad})$  is known as the prompt response function of the setup (PRD) that depends on the energy of the gamma rays in the cascade of interest and consequently the position of the centroids for such a cascade having lifetime much less compared to the time precision of the setup. In the present work involving Ge detectors (time precision in the order of few nanoseconds), (i) the cascades in $^{149}$Pm with known half life (T$_{1/2}$) of 2.53(3)~ns (156 - 115~keV, 246-115~keV, 301-115~keV and 401-115~keV)~\cite{nds_pm149}; (ii) the cascades in $^{150}$Sm (decay product of $^{150}$Pm) with known T$_{1/2}$ of 48.4(11)~ps (832-334 and 712-334 keV)~\cite{nds_pm150}, were utilized to understand the energy dependent prompt time response of the setup. Two of these experimentally determined centroid differences are shown in Fig~\ref{fig_pr_decay1} and Fig~\ref{fig_pr_decay2} (panel c in each figure) along with the corresponding energy projections (panel a and b in each figure). The experimentally determined $\Delta$C values are indicated inside the panel (c) of these figures. The time precision obtained for these distributions come out to be about 2~ns that determines the lower limit for time measurement using relative centroid shift methods from the present data at low energy. These two figures explain the determination of PRD from the known lifetime values of a level and the measured centroid differences.
\begin{figure}[ht!]
\begin{center}
\includegraphics[width=\columnwidth]{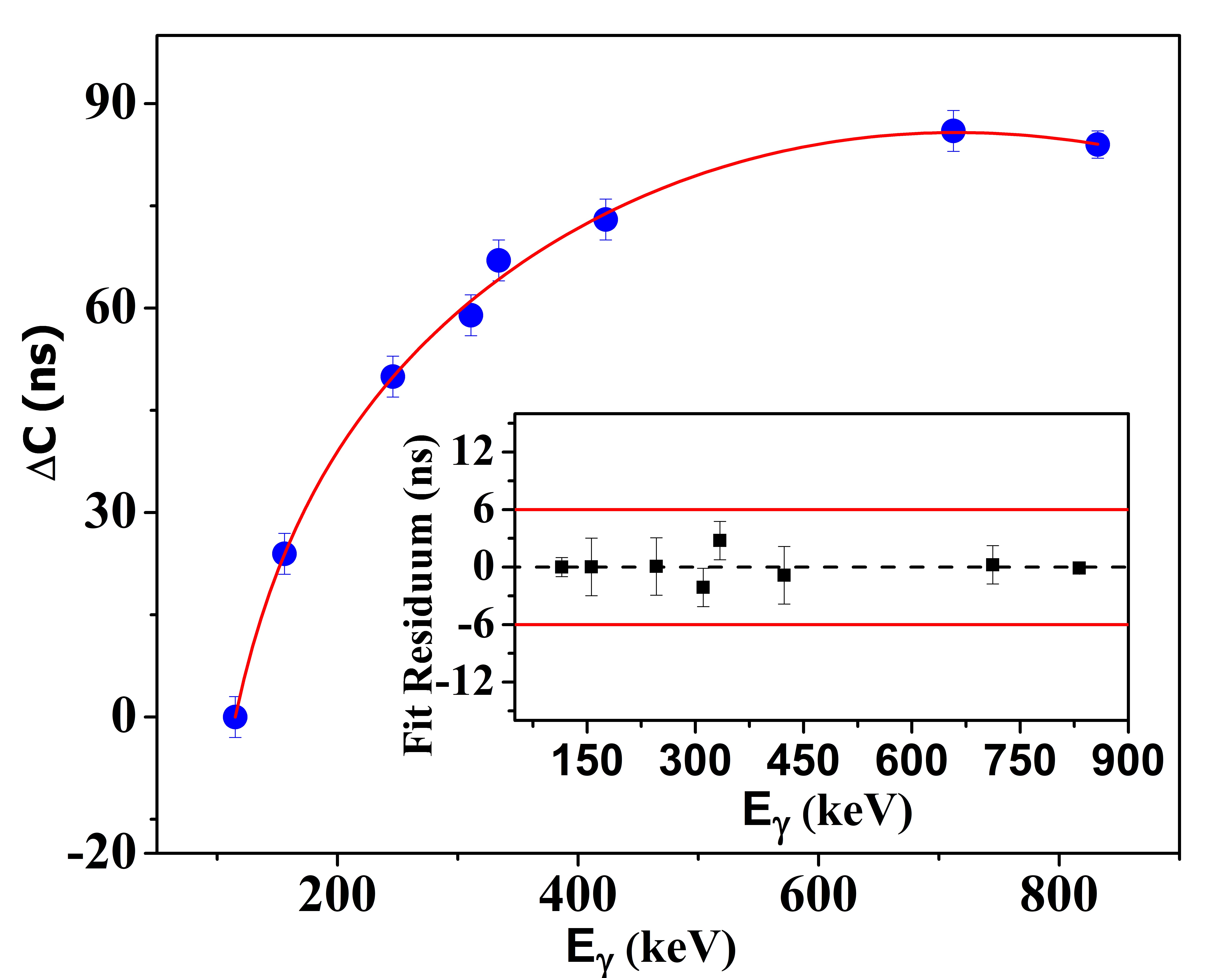}
\caption{(Color online) The PRD values (solid circles) obtained with the relative time difference analysis of several cascades are shown as a function of energy. The data points have been fitted with the relation described in text and the fit residuum of fit (solid squares) has been attached within the figure showing a 3$\sigma$ value of 6~ns.}
\label{prd}
\end{center}
\end{figure}

The measurement essentially requires a correction due to the underlying background in the gamma spectrum in the derived time information ($\delta$C). The details of such background correction procedure and its advantages over conventional background subtraction to derive time information~\cite{palit}, can be found in Ref.~\cite{regis2020}. Similar background corrections (t$_{corr}$), as utilized in our earlier works with fast scintillators~\cite{shefali,devesh,devesh2,safikul}, have been employed in the present work. The experimentally determined centroids, measured from the time distributions projected with selection of photopeaks from the energy projections of the detectors, include the time response of the underlying Compton background. Therefore, these centroids need to be corrected to determine the centroid difference corresponding to full energy peak. This is very important as Compton events at same energy will give completely different time information as compared to a full energy peak event. The $\Delta C_{BG}$ and t$_{corr}$ values were determined following the same set of relations given in equation 1 of Ref.~\cite{shefali}. The $\Delta C_{BG}$ values which were determined in reference to a particular photopeak energy, selecting energy gates on the background at different energy values surrounding the photopeak are also shown in Fig~\ref{fig_pr_decay1}, and Fig~\ref{fig_pr_decay2} (panel d and e in each figure) for the three cascades with known lifetimes. 
\begin{figure}[ht!]
\begin{center}
\includegraphics[width=\columnwidth]{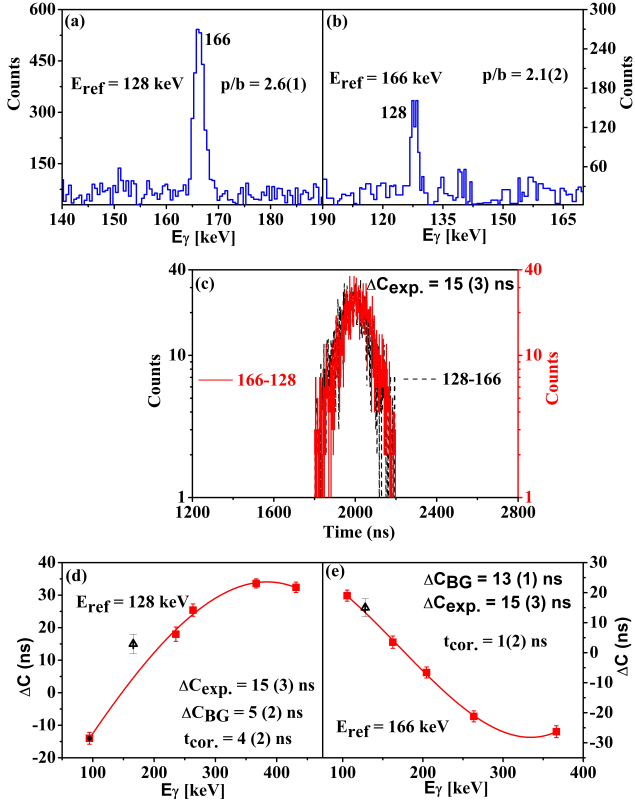}
\caption{(Color online) The lifetime analysis of 166-128~keV cascade of $^{150}$Pm has been explained. The gated projections for feeder and decay are shown in panels (a) and (b). The delayed (red solid line) and anti-delayed (black dashed line) projections for this cascade is displayed in panel (c) that provides $\Delta C_{expt}$=15(3)~ns (also shown with black triangle in (d) and (e)). The background corrections have been demonstrated in panel (d) and (e) (with red squares and solid lines), yielding at $\Delta C_{BG}(feeder)$ = 5(2)~ns, $\Delta C_{BG}(decay)$ = 13(1)~ns. With PRD (166,128) = 19(6) ,  we obtain a lifetime T$_{1/2}$ $< 2~ns$.}
\label{fig_decay1}
\end{center}
\end{figure}
\begin{figure}[ht!]
\begin{center}
\includegraphics[width=\columnwidth]{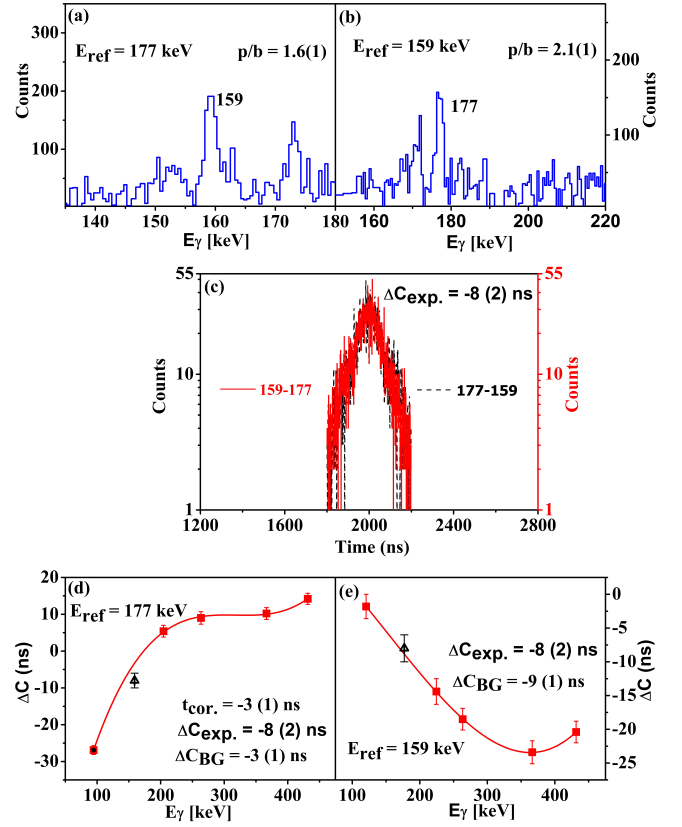}
\caption{(Color online) The lifetime analysis of 159-177~keV cascade of $^{150}$Pm has been explained. The gated projections for feeder and decay are shown in panels (a) and (b). The delayed (red solid line) and anti-delayed (black dashed line) projections for this cascade is displayed in panel (c) that provides $\Delta C_{expt}$=-8(2)~ns (also shown with black triangle in (d) and (e)). The background corrections have been demonstrated in panel (d) and (e) (with red squares and solid lines), yielding at $\Delta C_{BG}(feeder)$ = -3(1)~ns, $\Delta C_{BG}(decay)$ = -9(1)~ns yielding negligible correction for background. Comparing the $\Delta C_{FEP}$ = -10(3)~ns and the PRD value for the corresponding cascade we obtain a lifetime T$_{1/2}$ $< 2~ns$.}
\label{fig_decay2}
\end{center}
\end{figure}
\begin{figure}[ht!]
\begin{center}
\includegraphics[width=\columnwidth]{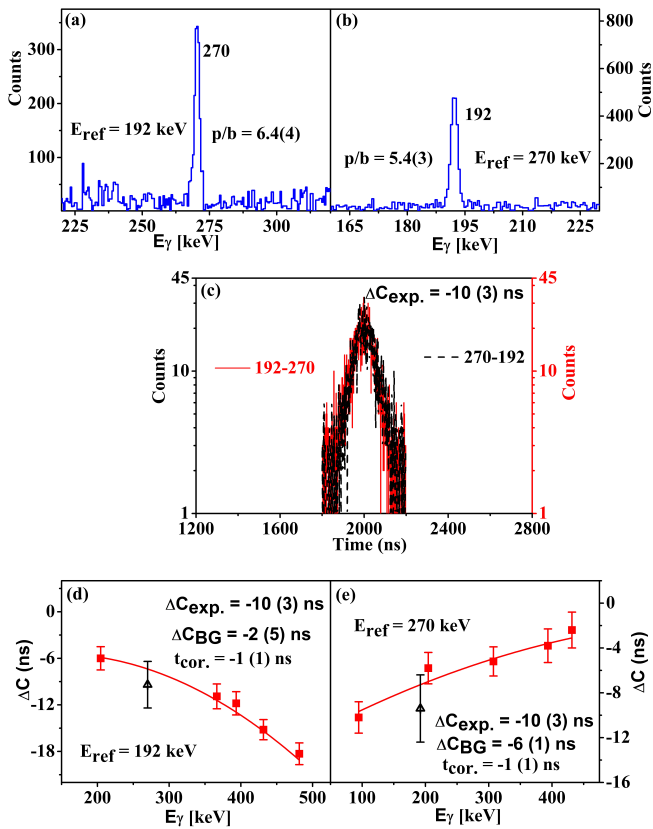}
\caption{(Color online) The lifetime analysis of 192-270~keV cascade of $^{149}$Pm has been explained. The gated projections for feeder and decay are shown in panels (a) and (b). The delayed (red solid line) and anti-delayed (black dashed line) projections for this cascade is displayed in panel (c) that provides $\Delta C_{expt}$=-10(3)~ns (also shown with black triangle in (d) and (e)). The background corrections have been demonstrated in panel (d) and (e) (with red squares and solid lines), yielding at $\Delta C_{BG}(decay)$ = -2(5)~ns,$\Delta C_{BG}(feeder)$ = -6(1)~ns . With PRD (192,270) = -18 (6) ns, we obtain a lifetime T$_{1/2}$ = $3(2)~ns$.}
\label{fig_decay3}
\end{center}
\end{figure}

The  $\Delta C_{BG}$ values at the particular energy of interest were determined from the fitted curve and were consequently utilized to determine the uncorrected background contribution ($\Delta C_{BG}$) at each energy of the cascade. The values were subsequently utilized to determine the correction factors t$_{corr}$ using peak to background ratios ({\it p/b}) using the set of equations shown in Ref.~\cite{shefali}. The centroid differences corresponding to full energy peak ($\Delta C_{FEP}$) were determined consequently before using these centroid differences to determine the PRD. The PRD values were subsequently obtained using the corrected centroid differences and known lifetime of a level.

The PRD values of all the cascades, so determined using the procedure described above and with respect to 115~keV, have been plotted in Fig.~\ref{prd} as a function of energy. The PRD values were fitted with the relation $PRD = \frac{a}{\sqrt{b+E_{\gamma}}} + c + dE_{\gamma} + eE_{\gamma}^2$, as elaborated in our earlier work~\cite{devesh}. This fitted curve represents the energy dependent prompt correction that is essential to be applied with the setups of Clover Ge detectors, especially in the low energy range, for measuring level lifetime through relative centroid difference method. This prompt calibration can be quite significant but often ignored assuming time response of the detectors to be same for both the transitions under consideration~\cite{ajay}. The fit residuum shown in Fig.~\ref{prd} represents the confidence limit obtained in the PRD. In the present work, 3$\sigma$ value of fit residuum (6~ns) was considered as PRD error.

The methodology described above has been used to determine $\Delta C_{FEP}$ values for the cascades of interest in $^{150}$Pm and also for the 197-270~keV cascade in $^{149}$Pm for which lifetime was known. Figure~\ref{fig_decay1}, Fig.~\ref{fig_decay2} and Fig.~\ref{fig_decay3} represent the lifetime analyses for 128~keV level and 177+X~keV level in $^{150}$Pm and 270~keV level in $^{149}$Pm,  respectively. The corresponding values of centroid differences, {\it p/b} and t$_{corr}$ are explained in the respective figures. It is observed that the present method reproduces the known lifetime in $^{149}$Pm quite well. The measured lifetime values in $^{150}$Pm for 128.3~keV and 177.2+X~keV levels come out to be less than 2~ns for both the cases. 
 
\section{Level by level discussion} 

The low lying level scheme of $^{150}$Pm developed in the present work, using the results of the data analysis described above, has been discussed in this section. In the present work, 16 new $\gamma$ rays have been placed in the level scheme compared to the existing data~\cite{bucu}. Placement of some of the $\gamma$ rays have been altered in the present work in comparison to that measurement. A detailed discussion of all the levels has been presented below. \\

{\it 1$^-$ ground state:}\\

The ground state of $^{150}$Pm has been assigned as having a spin parity of 1$^-$ following the discussions made in section~\ref{gs}. In the present work, all the $\gamma$ rays have been placed on this state maintaining the observed coincidence relationships among themselves. \\

{\it (2$^-$), 57.4 keV state:}\\

This 57.4~keV state was already known. As discussed in Section~\ref{57}, this level has been assigned with the spin parity of (2$^-$).\\

{\it 172.6, 249.2, 282.5, 319.8 \& 402.1~keV states:}\\

The 172.6~keV level is suggested from the coincidence of 57.4 keV and 115.2 keV $\gamma$ lines (Fig.~\ref{gates}(e)). One 172.6 keV $\gamma$ was found which is not in coincidence with the former two $\gamma$ lines but in coincidence with 76.6~keV, 109.9~keV,147.2~keV and 229.5~keV $\gamma$ lines. Accordingly, the 172.6 keV has been placed parallel to the 57.4 keV and 115.2 keV and thereby as decaying from the 172.6 keV level. The 109.9~keV $\gamma$ connects the previously known 282.5~keV (281.7~keV in~\cite{bucu}) level to the 172.6~keV level.

As the 76.6~keV, 109.9~keV, 147.2~keV and 229.5~keV $\gamma$ rays were found to be in coincidence with both 115.2~keV and 172.6~keV transitions, the 249.2~keV, 282.5~keV, 319.8~keV and 402.1~keV states are placed in the level scheme. Out of these, the 249.2~keV level was proposed from (d,$\alpha$) reaction but no $\gamma$ line was placed as decaying from this level in the earlier work. The placement of 76.6~keV $\gamma$ ray confirms the presence of this level. The 402.1~keV level is newly proposed in the present work. This is based on the placement of 229.5~keV $\gamma$ line which was observed in coincidence with neutron detector. 

We propose the assignments of (3$^-$) and (4), respectively, for the 172.6~keV and 319.8~keV levels. \\

{\it 128.3, 198.0, 226.7, 245.3, 251.2 \& 294.7~keV states:}\\

The excited state 128.3~keV was known (as 127.9~keV). The 198.0, 226.7, 245.3 and 294.7~keV states were also known which decay through the transitions of 69.7, 98.4, 117.0 and 166.4~keV $\gamma$ rays, respectively. In the present work, based on the coincidence of a new $\gamma$ ray of 122.9~keV, a 251.2~keV level was also placed in the level scheme (Fig.~\ref{gates}(b)). 

The spin of the 128.3~keV level was indicated as J = 2, following the angular distribution analysis of 128.3~keV transition. Spin parity assignments for the other levels in this group could not be made in the present work.

A lifetime of $ < 2~ns $ has been found for the 128.3~keV level as discussed in section~\ref{lt}.\\

{\it 130.3, 206.7, 220.8, 259.8, 283.1, 316.0 \& 381.8~keV states:}\\

The 130.3~keV level was known from earlier work and was also confirmed in the present work. Five transitions, viz. 76.4, 90.5, 129.5, 185.7 and 251.5~keV transitions were found to be in coincidence with the 130.3~keV $\gamma$ ray (Fig.~\ref{gates}(d)). Out of these transitions, only 90.5~keV transition was reported earlier and rest are newly observed ones. Among the new ones, the 76.4~keV shows coincidence with 76.4~keV itself and therefore, one of the 76.4~keV was placed as decaying from the 283.1~keV level which is already confirmed with the placement of 124.2~keV $\gamma$ ray in coincidence with 102.0~keV transition (Fig.~\ref{gates}(a)). Accordingly, the 206.7, 220.8, 259.8, 283.1, 316.0 and 381.8~keV states were placed in the level scheme of $^{150}$Pm. The 259.8~keV level decays to the 159.4~keV level via a 100.2~keV transition which was known from earlier work as well. In the present work, a 156.4 keV transition could be placed as decaying from the 316.0~keV level that also feeds the 159.4~keV level. The 251.5~keV transition has been placed in $^{150}$Pm as this $\gamma$ line shows coincidence with 130.3~keV (Fig.~\ref{gates}(d)). 

In the present work, no spin parity assignment was possible for these levels.\\

{\it 131.9, 208.9, 290.2 \& 370.7~keV states:}\\

One 75~keV $\gamma$ ray was tentatively known on a 172.3~keV level. However, in the present work, it is found that there are two transitions around 75~keV. Out of these two, the 74.5~keV shows coincidence with only 57.4~keV (Fig.~\ref{gates}(e)), thereby suggesting a level at 131.9~keV in $^{150}$Pm. The other transition of 76.6~keV was found to be in coincidence with both 115.2~keV and 172.6~keV, as discussed earlier. The 208.9, 290.2 and 370.6~keV levels were known earlier and also confirmed in the present work. 

In addition, the spin of J=3 has been proposed for the 208.9~keV state. \\

{\it 159.4, 240.4, 327.0, 331.0, 401.8 \& 416.5~keV states:}\\

In the present work, the 159.4~keV level was placed in the level scheme based on the coincidence of 102.0~keV $\gamma$ line with the 57.4~keV transition (Fig.~\ref{gates}(e)). However, no 159.4~keV was found to be decaying from this level, as was proposed earlier. On the contrary, one 159.2~keV $\gamma$ line could be placed on the 177.2+X~keV level, as discussed below. The $\gamma$ rays that were found in coincidence with 102.0~keV have been placed on the 159.4~keV level giving rise to the placements of 240.4, 327.0, 331.0, 401.8 and 416.5~keV states in the level scheme. Out of these, the 327.0~keV and 331.0~keV levels were already reported. One 81.0~keV transition was observed to be decaying from the newly proposed 240.4~keV level based on its coincidence with the 102.0~keV $\gamma$ line. This is, however, not the 81.3~keV placed earlier on the 208.9~keV. This is due to the fact that the 161.7~keV$\gamma$ ray placed on the 208.9~keV level was not found in coincidence with 102.0~keV, confirming the double placement of 81~keV (one 81.3~keV and the other 81.0~keV). 

In the present work, spin of J=2 is proposed for the 159.4~keV level. \\

{\it X, 177.2+X, 301.2+X, 336.4+X, 350.5+X, \& 454.8+X~keV states:}\\

All of these states are proposed to be placed on the 6$^-$ isomeric level (at X~keV) predicted from shell model calculation described in section~\ref{sm}. The states are placed based on the coincidence relationships of the 124.0, 159.2, 173.3 and 277.6~keV transitions with 177.2~keV (Fig.~\ref{gates}(c)). Out of these $\gamma$ rays, the 159.2~keV was also observed in the earlier work but was placed on the ground state. However, based on the observed coincidence of this transition (showing only 177.2~keV) and that of the 177.2~keV, the 159.2~keV is placed as decaying from a new 336.4~keV level. 

A lifetime of $ < 2~ns $ has been found for the 177.2+x~keV level as discussed in section~\ref{lt}.\\

\section{Discussion}
\begin{figure*}[htb]
\vspace{0cm}
\includegraphics[width=\textwidth]{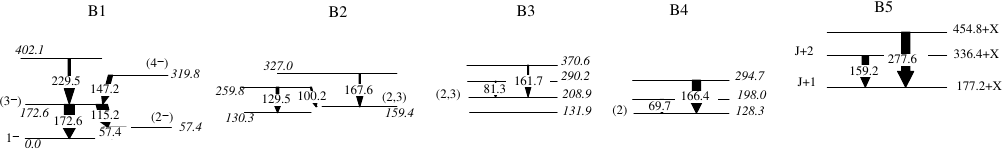} \caption{(Color online) Proposed band-like structures in $^{150}$Pm, inferred from the observed
level scheme (Fig.~\ref{level}) using the transitions that connect levels within a sequence.
The assignment of higher-lying members in each sequence remains tentative; the 131.9-keV
level is not linked to other levels in structure B3.}
\label{band}
\end{figure*}

The odd--odd nucleus $^{150}$Pm ($N=89$) lies in the transitional region of the nuclear chart and is expected to exhibit coexistence of near-spherical and collective structures with two competing potential minima~\cite{casten}. The quadrupole deformation parameter is reported to be $\beta_2=0.208$~\cite{def1}, suggesting that rotational correlations and quasiparticle coupling may play a role in its excitation spectrum. Although the experimentally established low-spin level scheme does not show an unambiguous rotational pattern, the available $\gamma$-ray decay information (Fig.~\ref{level}) indicates a few sequences that resemble band-like structures. These tentative sequences are proposed in Fig.~\ref{band}, guided
primarily by the observed intrasequence $\gamma$ decays and the predicted spin--parity of the associated levels. The ground-state spin--parity is taken as $1^-$, consistent with both the spherical shell-model (Sec.~\ref{sm}), projected shell-model (Sec.~\ref{psm}) and phenomenological (Sec.~\ref{pi}) predictions discussed below. The levels grouped in B4 and B5 of Fig.~\ref{band} are
assigned positive parity based on the conjectured parity-changing character of the 128.3- and 177.2-keV $\gamma$ transitions (see Sec.~\ref{intensity}). 

The observed level structure of odd--odd $^{150}$Pm is interpreted using complementary spherical shell-model and projected shell-model calculations along with phenomenology from the neighboring systematics. The spherical shell-model calculation is employed primarily to constrain the ground-state spin--parity and the dominant microscopic proton--neutron configurations of the lowest-lying states. The projected shell model, on the other hand, provides an intrinsic quasiparticle description built on a deformed mean field and is used to assess whether the proposed sequences can be understood as quasiparticle bands associated with a non-zero deformation, guided by the deformation of the even--even neighbor $^{150}$Nd.

\subsection{Interpretation with Shell Model calculation}
\label{sm}
 
Large-basis shell-model (LBSM) calculations were performed using the code NuShellX@MSU~\cite{nushell}, assuming $^{132}$Sn as an inert core. Proton valence particles were allowed to occupy the $50$--$82$ shell consisting of the $(1g_{7/2},\,2d_{5/2},\,2d_{3/2},\,3s_{1/2},\,1h_{11/2})$ orbitals, while neutrons were placed in the $82$--$126$ shell consisting of
$(1h_{9/2},\,2f_{7/2},\,2f_{5/2},\,3p_{3/2},\,3p_{1/2},\,1i_{13/2})$. The \textit{cwg} interaction~\cite{cwg} was employed; the single-particle energy of the proton $d_{5/2}$ orbital was kept unchanged as in Ref.~\cite{shergur}.

These calculations are microscopic in nature and often provide robust predictions for the ground-state spin--parity and dominant configurations~\cite{ce139,la137}. However, absolute excitation energies, level densities, and the placement of collective structures can be sensitive to the adopted
model-space truncations. In the present work, truncations were introduced in both proton and neutron spaces to keep the calculation feasible. The truncation scheme was chosen such that the low-lying spectra of the neighbouring odd-$Z$ nucleus $^{149}$Pm, odd-$N$ nucleus $^{149}$Nd,
and odd--odd nucleus $^{148}$Pm are reasonably reproduced. This benchmarking is discussed below before presenting the results for $^{150}$Pm.

\subsubsection{Verification of calculation}
\label{ver}
\begin{figure*}[ht!]
\begin{center}
\includegraphics[width=\textwidth]{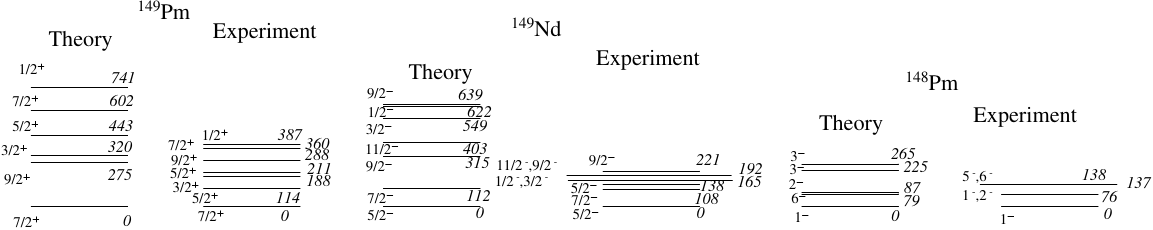}
\caption{(Color online) Low lying level structure of the neighbors of $^{150}$Pm obtained from shell model calculation are shown in comparison to the experimental data~\cite{nds_pm149,nds_pm148}. The calculation has been done using the particle restriction shown in Table~\ref{restrict}.}
\label{sm_neighbor}
\end{center}
\end{figure*}
\begin{table*}[ht!]
\begin{center}
\caption{The Particle restriction used in the calculation for the shell model calculation.}
\begin{tabular}{cccccccccccc}
\hline
Nucleus&\multicolumn{5}{c}{Proton}&\multicolumn{5}{c}{Neutron}\\
&$\pi g_{7/2}$&$\pi d_{5/2}$&$\pi h_{11/2}$&$\pi d_{3/2}$&$\pi s_{1/2}$&$\nu f_{7/2}$&$\nu h_{9/2}$&$\nu f_{5/2}$&$\nu p_{3/2}$&$\nu p_{1/2}$&$\nu i_{13/2}$\\
\hline
\hline
$^{149}$Pm&( 7, 8 )&( 2, 4 )&( 0, 0 )&( 0, 2 )&( 0, 2 )&( 3, 6 )& ( 0 , 1 )&( 0 , 0 )&( 0, 3 )&( 0 , 2 )&( 0 , 0 )\\
$^{149}$Nd&( 6, 8 )&( 2, 6 )&( 0, 0 )&( 0, 1 )&( 0, 1 )&( 5, 7 )& ( 2 , 7 )&( 0 , 0 )&( 0, 0 )&( 0 , 0 )&( 0 , 0 )\\
$^{148}$Pm&( 7, 8 )&( 2, 4 )&( 0, 0 )&( 0, 2 )&( 0, 2 )&( 3, 6 )& ( 0 , 1 )&( 0 , 0 )&( 0, 3 )&( 0 , 2 )&( 0 , 0 )\\
$^{150}$Pm&( 7, 8 )&( 2, 4 )&( 0, 0 )&( 0, 2 )&( 0, 2 )&( 5, 7 )& ( 2 , 7 )&( 0 , 0 )&( 0, 0 )&( 0 , 0 )&( 0 , 0 )\\
\hline
\hline
\end{tabular}
\label{restrict}
\end{center}
\end{table*}
\begin{table*}[ht!]
\centering
\caption{Shell model configurations for low-lying states in $^{148}$Pm and $^{150}$Pm.}
\label{conf}
\begin{tabular}{cccccccccl}
\hline\hline
\multirow{2}{*}{Nucleus} & \multicolumn{2}{c}{$E_x$ (keV)} & \multicolumn{2}{c}{$J^\pi$} & $I_\pi$ & $I_\nu$ & \% prob. & Major partition & \% prob. \\
 & expt. & calc. & expt. & calc. & & & (a) & & (b) \\
\hline\hline 
\multirow{10}{*}{$^{148}$Pm} 
 & 0 & 0 & $1^-$ & $1^-$ & $5/2$ & $7/2$ & 58 & $\pi(1g_{7/2}^7 2d_{5/2}^4) \otimes \nu(2f_{7/2}^{5})$ & 20 \\
 & & & & & & &  & $\pi(1g_{7/2}^7 2d_{5/2}^4) \otimes \nu(2f_{7/2}^{4}2p_{3/2}^{1})$ & 12 \\
 & & & & & $7/2$ & $7/2$ & 11 & $\pi(1g_{7/2}^7 2d_{5/2}^4) \otimes \nu(2f_{7/2}^{5})$ & 4 \\
 & & & & & $7/2$ & $9/2$ & 16 & $\pi(1g_{7/2}^7 2d_{5/2}^4) \otimes \nu(2f_{7/2}^{5})$ & 4 \\
\cline{2-10}
 & 76 & 87 & $1^-,2^-$ & $2^-$ & $5/2$ & $7/2$ & 37 & $\pi(1g_{7/2}^7 2d_{5/2}^4) \otimes \nu(2f_{7/2}^{5})$ & 13 \\
 & & & & & & & & $\pi(1g_{7/2}^7 2d_{5/2}^4) \otimes \nu(2f_{7/2}^{4}2p_{3/2}^{1})$ & 7 \\
 & & & & & $7/2$ & $7/2$ & 23 & $\pi(1g_{7/2}^7 2d_{5/2}^4) \otimes \nu(2f_{7/2}^{5})$ & 8 \\
\cline{2-10}
 & 138 & 79 & $5^-,6^-$ & $6^-$ & $5/2$ & $7/2$ & 41 & $\pi(1g_{7/2}^7 2d_{5/2}^4) \otimes \nu(2f_{7/2}^{5})$ & 15 \\
 & & & & & & & & $\pi(1g_{7/2}^7 2d_{5/2}^4) \otimes \nu(2f_{7/2}^{4}2p_{3/2}^{1})$ & 8 \\
 & & & & & $7/2$ & $7/2$ & 19 & $\pi(1g_{7/2}^7 2d_{5/2}^4) \otimes \nu(2f_{7/2}^{5})$ & 6 \\
\hline
\multirow{10}{*}{$^{150}$Pm}
 & 0 & 0 & $1^-$ & $1^-$ & $5/2$ & $7/2$ & 52 & $\pi(1g_{7/2}^7 2d_{5/2}^4) \otimes \nu(1h_{9/2}^{2}2f_{7/2}^{5})$ & 33 \\
 & & & & & & & & $\pi(1g_{7/2}^7 2d_{5/2}^{3}2d_{3/2}^{1}) \otimes \nu(1h_{9/2}^{2}2f_{7/2}^{5})$ & 6 \\
 & & & & & & & & $\pi(1g_{7/2}^7 2d_{5/2}^{2}2d_{3/2}^{2}) \otimes \nu(1h_{9/2}^{2}2f_{7/2}^{5})$ & 5 \\
 & & & & & & & & $\pi(1g_{7/2}^7 2d_{5/2}^{3}3s_{1/2}^{1}) \otimes \nu(1h_{9/2}^{2}2f_{7/2}^{5})$ & 5 \\
 & & & & & $7/2$ & $7/2$ & 13 & $\pi(1g_{7/2}^7 2d_{5/2}^4) \otimes \nu(1h_{9/2}^{2}2f_{7/2}^{5})$ & 8 \\
 & & & & & $7/2$ & $9/2$ & 15 & $\pi(1g_{7/2}^7 2d_{5/2}^4) \otimes \nu(1h_{9/2}^{2}2f_{7/2}^{5})$ & 7 \\
\cline{2-10}
 &  & 7 &  & $6^-$ & $5/2$ & $7/2$ & 36 & $\pi(1g_{7/2}^7 2d_{5/2}^4) \otimes \nu(1h_{9/2}^{2}2f_{7/2}^{5})$ & 23 \\
 & & & & & $7/2$ & $7/2$ & 22 & $\pi(1g_{7/2}^7 2d_{5/2}^4) \otimes \nu(1h_{9/2}^{2}2f_{7/2}^{5})$ & 14 \\
\cline{2-10}
 & 57 & 31 & $(2^-)$ & $2^-$ & $5/2$ & $7/2$ & 36 & $\pi(1g_{7/2}^7 2d_{5/2}^4) \otimes \nu(1h_{9/2}^{2}2f_{7/2}^{5})$ & 23 \\
 & & & & & $7/2$ & $7/2$ & 25 & $\pi(1g_{7/2}^7 2d_{5/2}^4) \otimes \nu(1h_{9/2}^{2}2f_{7/2}^{5})$ & 16 \\
 \hline\hline
\bottomrule
\end{tabular}
\end{table*}
\begin{table*}[ht!]
\begin{center}
\caption{Experimental and shell model transition probabilities for selected transitions in $^{149}$Pm and $^{149}$Nd.}
\label{trans_prob}
\begin{tabular}{cccccc}
\hline
\textbf{Nucleus} & \textbf{$I_i \rightarrow I_f$} & \multicolumn{2}{c}{\textbf{B(E2) [W.u.]}} & \multicolumn{2}{c}{\textbf{B(M1) [W.u.]}} \\
& & Expt.~\cite{nds_pm149}& SM & Expt.~\cite{nds_pm149} & SM \\
\hline
$^{149}$Pm & $3/2^+ \rightarrow 7/2^+$ & $3.2^{+0.6}_{-0.5}$ & 4.8 & -- & -- \\
           & $5/2^+ \rightarrow 7/2^+$ & $3.0^{+0.8}_{-0.7}$ & 1.0 & $(2.751^{+0.041}_{-0.043}) \times 10^{-3}$ & $2.715 \times 10^{-3}$ \\
\hline
$^{149}$Nd & $7/2^- \rightarrow 5/2^-$ & - & - & $0.043^{+0.031}_{-0.013}$ & 0.120 \\        
\hline
\end{tabular}
\end{center}
\end{table*}
The calculated low-lying level schemes of the neighbouring nuclei $^{149}$Pm, $^{149}$Nd, and $^{148}$Pm are compared with the experimental data~\cite{nds_pm149,nds_pm148} in Fig.~\ref{sm_neighbor}. The particle restrictions adopted for these nuclei are summarized in Table~\ref{restrict}.

For all three benchmark nuclei, the ground state is correctly reproduced and the overall pattern of the lowest-lying states is described reasonably well, although some interchange of nearby levels is observed. Such deviations are likely a consequence of the truncations rather
than an indication of qualitatively incorrect dominant configurations. The calculated wavefunctions indicate that the lowest states are dominated by a single unpaired proton in the $g_{7/2}$ orbital for $^{149}$Pm and a single unpaired neutron in the $f_{7/2}$ orbital for $^{149}$Nd, consistent with expectations in this mass region.

To further test the reliability of the calculated wavefunctions in the benchmark nuclei, electromagnetic transition probabilities among selected low-lying levels were calculated for $^{149}$Pm and $^{149}$Nd and compared with the evaluated experimental values. As shown in
Table~\ref{trans_prob}, the shell-model results reproduce the measured transition strengths within reasonable agreement. This supports the conclusion that, despite the imposed truncations, the calculation captures the dominant structure of the lowest-lying states relevant for the present discussion.

On successful verification of the validity of the restricted calculations for the neighboring odd-A nuclei, the shell model calculation was performed for odd-odd $^{148}$Pm for which few low lying states are known from experiment. In this case, the particle restriction (Table~\ref{restrict}) for proton was same as that used for $^{149}$Pm (Z = 61) and that for neutron was similar to the one used for $^{149}$Nd (N = 89). The comparison of experiment and calculation for $^{148}$Pm are also shown in Fig.~\ref{sm_neighbor} along with the odd-A ones. It is observed that the lowest three energy levels are reproduced quite well and the corresponding wavefunctions are given in Table~\ref{conf}. The level structure of $^{150}$Pm was, therefore, calculated using same model space, interaction and particle restrictions as used for $^{148}$Pm and the corresponding results are discussed in the following section in comparison to the experimental data.

\subsubsection{Levels in $^{150}$Pm}

\begin{figure}[ht!]
\begin{center}
\includegraphics[width=\columnwidth]{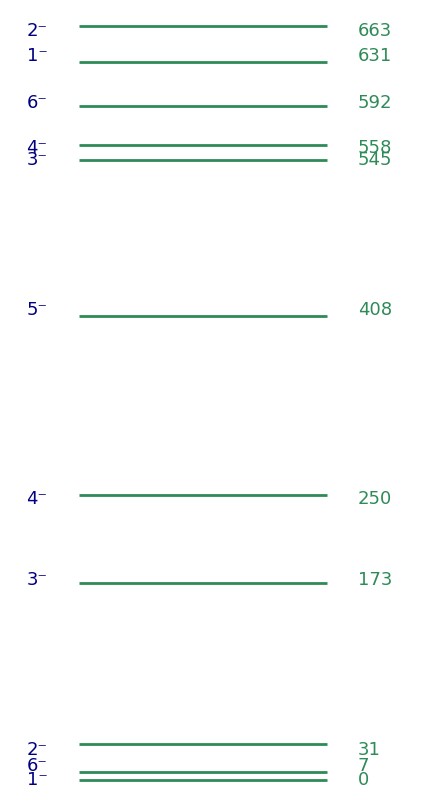}
\caption{The negative parity levels in $^{150}$Pm, calculated with shell model using NUSHELLX. Energy levels up to J$^{\pi} = 6^-$ and 663~keV are shown. The level density found from calculation is significantly lower compared to the observed levels in experiment below 500~keV. The restrictions applied to the shell model calculation could be responsible for predicted higher excitations and lower level densities for the higher lying levels.}
\label{150Pm_theory}
\end{center}
\end{figure}

The present shell-model calculation predicts a $1^-$ ground state for $^{150}$Pm and a closely lying $2^-$ state at 31~keV, consistent with the experimental ground-state assignment of Hoshi \textit{et al.}~\cite{hoshi}. In addition, a low-lying $6^-$ state is predicted at 7~keV, which is not inconsistent with the systematics of isomeric states in this region~\cite{pmbeta}, although its experimental placement and decay remain uncertain. The dominant configurations of these low-lying states are summarized in Table~\ref{conf} in comparison to those in $^{148}$Pm. Overall, the calculated $1^-$, $2^-$, and
$6^-$ states emerge primarily from coupling the leading proton and neutron configurations in the truncated space, producing a set of closely spaced negative-parity states at low excitation.

Additional negative-parity states obtained in the calculation are shown in Fig.~\ref{150Pm_theory}. The calculated level density below $\sim$500~keV is significantly smaller than that suggested by the experimental level scheme, likely reflecting the limitations of the truncated configuration space. Furthermore, positive-parity states require
excitations involving $\pi h_{11/2}$ and/or $\nu i_{13/2}$ and are predicted above $\sim$1~MeV within the present truncations. This makes it difficult for the restricted spherical shell model alone to address the possible band-like positive-parity sequences proposed experimentally. In particular, the low-lying $1^+$ state reported at $\sim$110~keV~\cite{gauss,bucu} (not observed in the present work) may involve correlations beyond those captured in the present truncated shell-model space, and could therefore have a more collective component. Consistent with this possibility, interacting boson--fermion--fermion model (IBFFM) calculations~\cite{nomura} predict several low-lying positive-parity states in $^{150}$Pm while retaining a $1^-$ ground state.

In summary, the restricted spherical shell-model calculation is used here primarily to constrain the ground-state spin--parity and to identify the dominant microscopic proton--neutron configurations for the lowest-lying negative-parity states. However, the calculated low-energy
level density and the placement of positive-parity excitations are sensitive to the imposed truncations and therefore do not allow a reliable assessment of deformation effects or rotational correlations. Given that $^{150}$Pm lies in a transitional region and has nonzero quadrupole deformation, we therefore employ the projected shell model in the next section. The PSM provides an intrinsic quasiparticle description built on a deformed mean field and incorporates rotational correlations through angular-momentum projection, making it well suited to examine the role of deformation and quasiparticle coupling in the low-lying structure of $^{150}$Pm.

\subsection{Interpretation with Projected Shell model calculation}
\label{psm}

To provide an insight into the nature of the band structures observed in odd-odd  $^{150}$Pm, numerical calculations have been performed using the
microscopic projected shell model (PSM)  approach \cite{Hara1995}.
The PSM represents a beyond the mean-field framework that combines the mean-field and conventional shell model approach by employing the angular-momentum
projected deformed Nilsson states as the basis configurations in the shell model diagonalization \cite{Hara1995}. This approach begins with a
deformed mean field, which provides an optimum  starting point for describing nuclei with significant deformations, and
introduces essential pairing correlations through the quasi-particle BCS transformation. The model employs a pairing plus quadrupole-quadrupole Hamiltonian and is given by:
\begin{equation}
\hat{H}=\hat{H}_{0}-\frac{1}{2}\chi\sum_{\mu}\hat{Q}_{\mu}^{\dagger}\hat{Q}_{\mu}-G_{M}\hat{P}^{\dagger}\hat{P}-G_{Q}\sum_{\mu}\hat{P}_{\mu}^{\dagger}\hat{P}_{\mu},
\end{equation}
where the first term $\hat{H}_{0}$ represents the harmonic-oscillator single-particle Hamiltonian with proper spin-orbit coupling based on the Nilsson potential \cite{Nilsson1969}, while the subsequent terms are the quadrupole-quadrupole, monopole pairing, and quadrupole pairing interactions. The interaction strengths are determined through physical considerations : the quadrupole force strength $\chi$ is adjusted self-consistently to the known or expected deformation of the
system, while the pairing strengths are derived from established formulas \cite{Dieterich1975} and fine-tuned to reproduce experimental odd-even mass differences.
The quadrupole pairing strength is typically set at approximately one-fifth of the monopole pairing constant.

The next stage in the PSM approach involves diagonalizing the Hamiltonian in a basis of angular-momentum-projected multi-quasiparticle states. The configuration space employed in the present study consists of one quasiproton coupled to one quasineutron i.e, $ \hat P^I_{MK}~a^\dagger_{\nu}~a^\dagger_{\pi}|\Phi\rangle$, where $\hat{P}^{I}_{MK}$ is the angular momentum projection operator that projects the quantum numbers $I$, $K$ and $M$.
\begin{figure}[htb]
\vspace{0cm}
\includegraphics[width = \columnwidth]{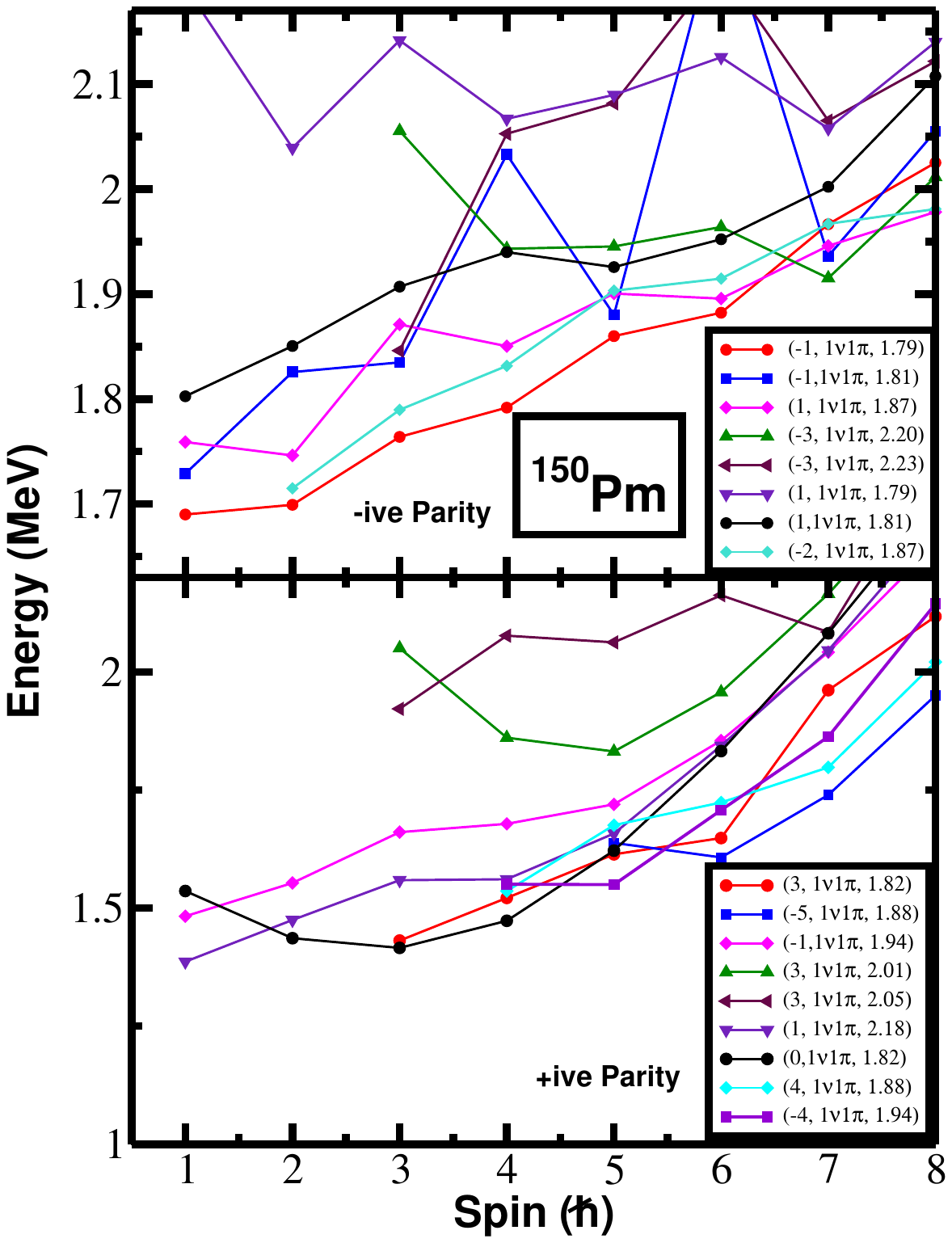} \caption{(Color
online)  Projected energies are shown before diagonalization of the
shell model Hamiltonian for $^{150}$Pm. The bands are labelled by
three quantities : K-quantum number, group structure, energy of the quasiparticle
state. For instance, $(3, 1\nu 1\pi,1.82)$ designates two-quasiparticle state having intrinsic energy of 1.82 MeV and K$=3$. } \label{150Pm_BandDiagram.pdf}
\end{figure}
\begin{figure*}[htb]
\vspace{0cm}
\includegraphics[width=\textwidth]{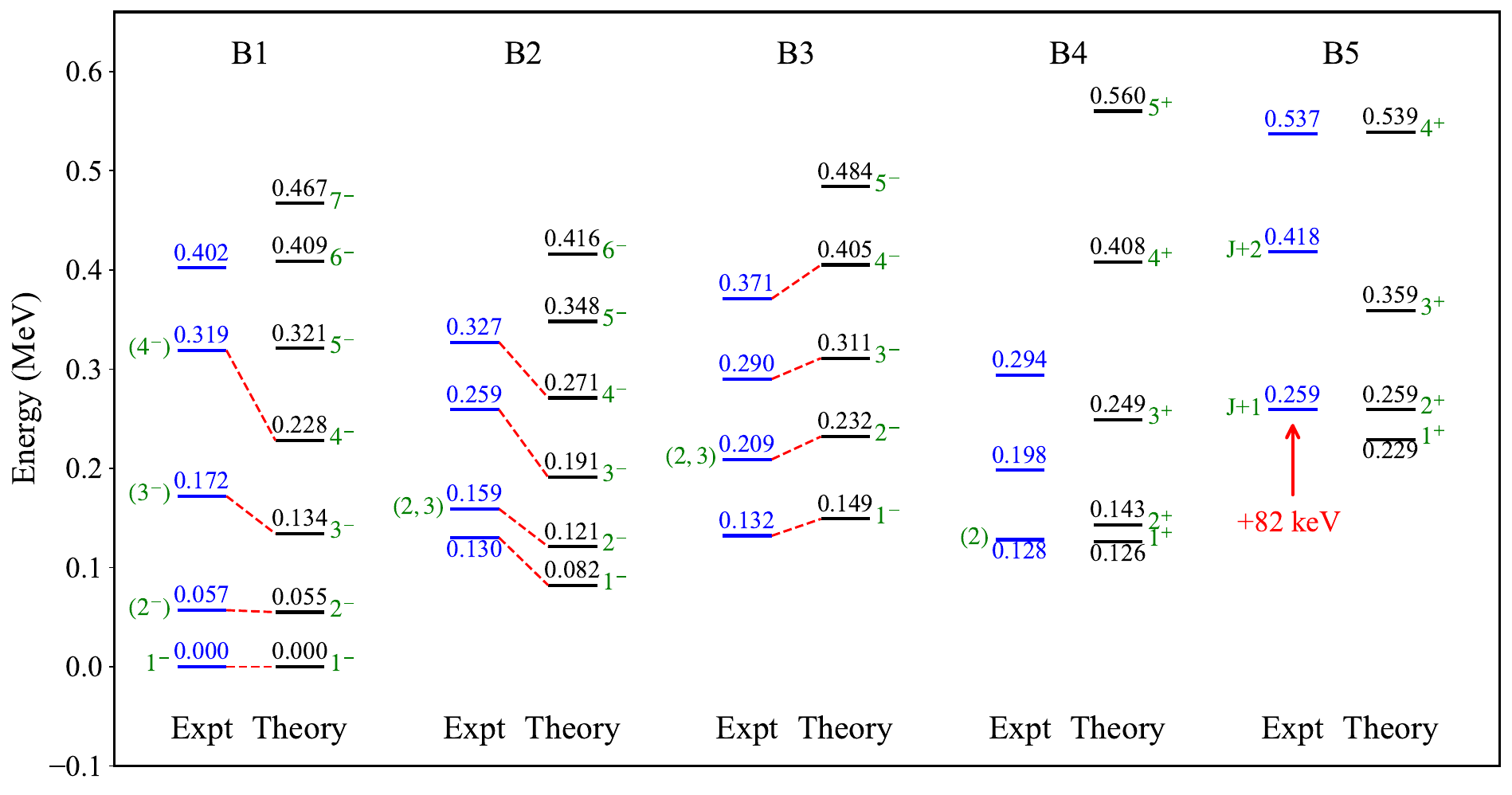} \caption{(Color
online)  Comparison of experimentally observed level energies of negative parity bands B1, B2, and B3,  and positive parity bands B4 and B5 with the values obtained from PSM calculations. The experimental energies of band B5 has been shown by adjusting the energy of the 177+x~keV levels with the calculated energy of the 2$^+$ level of this band, considering x = 82~keV
.} \label{Figure_1.pdf}
\end{figure*}
The PSM has proven particularly useful for $A\sim150$ nuclei, as demonstrated in previous studies \cite{Wang2018,Velazquez1999,Sun1996}, providing a reliable description of rotational bands and transition probabilities in this mass region.
The present calculations employ specific input parameters, the quadrupole deformation parameter
$\epsilon \sim 0.208$, which is consistent with values reported in Ref. \cite{Moller1995}.
The pairing interaction parameters are set as $G_1 = 21.00$ MeV and $G_2 = 10.70$ MeV, with the quadrupole pairing strength $G_Q$ set to 0.16 times the monopole pairing constant $G_M$. The Nilsson parameters $\kappa$ and $\mu$ are adopted from Ref. \cite{Nilsson1969}, and the single-particle configurations include three major shells: $N=3,4,5$, both for protons and neutrons. To generate
the positive parity states in  $^{150}$Pm, the  valence neutrons and protons are occupying the N= 5 shell, and for negative
parity valence neutron (proton) are in the N= 5 (4) shell. 

We shall first discuss the PSM results before the shell model diagonalization. In this first stage of the calculation, the projection is performed for each intrinsic Nilsson state and  the projected energy as a function angular momentum is plotted. This plot
referred to as the band diagram, depicted in  Fig. \ref{150Pm_BandDiagram.pdf}, is quite instructive as it provides
information on the intrinsic configuration of the band structure.
It is noted from Fig. \ref{150Pm_BandDiagram.pdf} (upper panel) that the negative parity lowest band originates
from the $1\nu 1\pi$ configuration with quantum number $K=-1$, and quasiparticle energy of 1.79 MeV, labelled
as $(-1, 1\nu1\pi, 1.79)$ in the  Fig. \ref{150Pm_BandDiagram.pdf}. This band is crossed at spin I=7 by another two-quasiparticle  band having a
different intrinsic configuration of $(-3, 1\nu1\pi, 2.20)$. The
first excited band originates from $1\nu 1\pi$ configuration with  $K=-2$ and quasiparticle energy of 1.87 MeV.
As is evident from Fig.~\ref{150Pm_BandDiagram.pdf}, the second excited band depicts crossings with several other
configurations consisting of $(1, 1\nu1\pi, 1.87)$ and $(-1, 1\nu1\pi, 1.81)$.
In the positive  parity band diagram, lower panel of  Fig. \ref{150Pm_BandDiagram.pdf}, The lowest
band structure originates from the two-quasiparticle state, $1\nu 1\pi$ with competing configurations of $(0, 1\nu 1\pi, 1.82)$ and $(-5, 1\nu1\pi, 1.88)$.

The projected bands shown in Fig.~\ref{150Pm_BandDiagram.pdf} and many more around the Fermi surface are employed
to diagonalize the shell model Hamiltonian consisting of pairing and quadrupole-quadrupole interaction terms.
Following diagonalization, the resulting lowest three negative-parity bands and the lowest two positive-parity band are
presented in Fig.~\ref{Figure_1.pdf}. It is quite evident from the  Fig.  \ref{Figure_1.pdf}  that the negative parity band structures dominate the low-energy spectrum with Band B1 emerging as the yrast sequence. 
It is quite evident from Fig.~\ref{Figure_1.pdf} that the  comparison between theoretical and experimental band structures shows a reasonable
agreement with some deviations noted in the high-spin region.
\begin{table*}
\centering
\small 
\tabcolsep 0.09cm 
\caption{Wavefunctions amplitudes $g^{\sigma I}_{K\kappa} $ from PSM calculations for lowest spins of $^{150}$Pm.}
\begin{tabular}{cccc}
\toprule
\hline\hline
{Spin}  & {(K, Str, $E_{qp}$) $g_1^2$} & {(K, Str, $E_{qp}$) $g_2^2$} & {(K, Str, $E_{qp}$) $g_3^2$}\\
\hline
\hline
\multicolumn{4}{c}{{Band B1}} \\
\hline
1$^{-}$   & (-1, 1$\nu$1$\pi$, 1.79) 0.800  & (-1, 1$\nu$1$\pi$, 1.81) 0.120 & (1, 1$\nu$1$\pi$, 1.87) 0.04\\
2$^{-}$   & (-1, 1$\nu$1$\pi$, 1.79) 0.655 & (-2, 1$\nu$1$\pi$, 1.87) 0.203 & (1, 1$\nu$1$\pi$, 1.87) 0.07\\
3$^{-}$   & (-1, 1$\nu$1$\pi$, 1.79) 0.621 & (-2, 1$\nu$1$\pi$, 1.87) 0.210 & (1, 1$\nu$1$\pi$, 1.87) 0.021\\
4$^{-}$   & (-1, 1$\nu$1$\pi$, 1.79) 0.571 & (-2, 1$\nu$1$\pi$, 1.87) 0.193 & (1, 1$\nu$1$\pi$, 1.87) 0.04\\
5$^{-}$   & (-1, 1$\nu$1$\pi$, 1.79) 0.210 & (-2, 1$\nu$1$\pi$, 1.87) 0.131 & (-1, 1$\nu$1$\pi$, 1.81) 0.46\\
\hline
\multicolumn{4}{c}{{Band B2}} \\
\hline
1$^{-}$   & (-1, 1$\nu$1$\pi$, 1.81) 0.510  & (1, 1$\nu$1$\pi$, 1.81) 0.01 & (1, 1$\nu$1$\pi$, 1.87) 0.30\\
2$^{-}$   & (-2, 1$\nu$1$\pi$, 1.87) 0.550 & (1, 1$\nu$1$\pi$, 1.87) 0.201 & (-1, 1$\nu$1$\pi$, 1.81) 0.01\\
3$^{-}$   & (-2, 1$\nu$1$\pi$, 1.87) 0.510 & (1, 1$\nu$1$\pi$, 1.87) 0.009 & (-1, 1$\nu$1$\pi$, 1.81) 0.29\\
4$^{-}$   & (-2, 1$\nu$1$\pi$, 1.87) 0.472 & (1, 1$\nu$1$\pi$, 1.87) 0.172 & (-1, 1$\nu$1$\pi$, 1.81) 0.03\\
\hline
\multicolumn{4}{c}{{Band B3}} \\
\hline
1$^{-}$   & (1, 1$\nu$1$\pi$, 1.87) 0.501  & (1, 1$\nu$1$\pi$, 1.81) 0.123 & (-1, 1$\nu$1$\pi$, 1.81) 0.134\\
2$^{-}$   & (1, 1$\nu$1$\pi$, 1.87) 0.121 & (1, 1$\nu$1$\pi$, 1.81) 0.022 & (-1, 1$\nu$1$\pi$, 1.81) 0.523\\
3$^{-}$   & (1, 1$\nu$1$\pi$, 1.87) 0.452 & (1, 1$\nu$1$\pi$, 1.81) 0.009 & (-1, 1$\nu$1$\pi$, 1.81) 0.372\\
4$^{-}$   & (1, 1$\nu$1$\pi$, 1.87) 0.210 & (1, 1$\nu$1$\pi$, 1.81) 0.001 & (-1, 1$\nu$1$\pi$, 1.81) 0.410\\
\hline
\multicolumn{4}{c}{{Band B4}} \\
\hline
1$^{+}$   & (1, 1$\nu$1$\pi$, 2.18) 0.790  & (-1, 1$\nu$1$\pi$, 1.94) 0.110 & (0, 1$\nu$1$\pi$, 1.82) 0.001\\
2$^{+}$   & (1, 1$\nu$1$\pi$, 2.18) 0.221  & (-1, 1$\nu$1$\pi$, 1.94) 0.013 & (0, 1$\nu$1$\pi$, 1.82) 0.522\\
3$^{+}$   & (1, 1$\nu$1$\pi$, 2.18) 0.052 &  (0, 1$\nu$1$\pi$, 1.82) 0.451 & (3, 1$\nu$1$\pi$, 1.82) 0.252\\
4$^{+}$   & (4, 1$\nu$1$\pi$, 1.88) 0.110 &  (0, 1$\nu$1$\pi$, 1.82) 0.331 & (3, 1$\nu$1$\pi$, 1.82) 0.210\\
\hline
\multicolumn{4}{c}{{Band B5}} \\
\hline
1$^{+}$   & (-1, 1$\nu$1$\pi$, 1.94) 0.632   & (1, 1$\nu$1$\pi$, 2.18) 0.283  & - \\
2$^{+}$   & (-1, 1$\nu$1$\pi$, 1.94) 0.201  & (1, 1$\nu$1$\pi$, 2.18) 0.501 & (0, 1$\nu$1$\pi$, 1.82) 0.131\\
3$^{+}$   & (-1, 1$\nu$1$\pi$, 1.94) 0.102 &  (0, 1$\nu$1$\pi$, 1.82) 0.212 & (1, 1$\nu$1$\pi$, 1.82) 0.321\\
4$^{+}$   & (4, 1$\nu$1$\pi$, 1.88) 0.237 &  (0, 1$\nu$1$\pi$, 1.82) 0.242 & (3, 1$\nu$1$\pi$, 1.82) 0.335\\
\hline \hline
\end{tabular}
\label{gi2}
\end{table*}
\begin{table*}
\small 
\centering
\tabcolsep 0.16cm 
\caption{Nilsson wave-function amplitudes  from PSM calculations for $^{150}$Pm}
\begin{tabular}{c c c c c c c c c}
\hline  \hline
  (K, Str, $E_{qp}$) &  & $E_{sp}$  &  $E_{Nilsson}$& $K$ 
& $|a|^{2}$ \\
\hline 
 (-1, 1$\nu$1$\pi$, 1.79) & 1$\nu$ & 0.659 & 53.325 & $1/2$
& 0.768($2f_{7/2}$),\;0.137($1h_{9/2}$),\;0.036($3p_{3/2}$)\\
         & 1$\pi$ & 1.128 & 41.260 & $-3/2$
& 0.903($2d_{5/2}$),\;0.0749($1g_{7/2}$) \\
\hline 
(-1, 1$\nu$1$\pi$, 1.81)   & 1$\nu$ & 0.659 & 53.325 & $-3/2$
& 0.768($2f_{7/2}$),\;0.137($1h_{9/2}$),\;0.036 ($3p_{3/2}$) \\
         & 1$\pi$ & 1.154 & 41.764 & $1/2$
& 0.997($1g_{7/2}$),\;0.002($1g_{9/2}$) \\
\hline 
 (1, 1$\nu$1$\pi$, 1.87) & 1$\nu$ & 0.659 & 53.325 & $-3/2$
& 0.768($2f_{7/2}$),\;0.137($1h_{9/2}$),\;0.036($3p_{3/2}$)   \\
         & 1$\pi$ & 1.216 & 40.959 & $5/2$
& 0.962($1g_{7/2}$),\;0.033($2d_{5/2}$) \\
\hline \hline 
 (3, 1$\nu$1$\pi$, 1.82) &1$\nu$  &0.501  & 53.491 & $1/2$
& 0.880($1h_{9/2}$),\;0.080($2f_{5/2}$)  \\
         & 1$\pi$      &  1.319      & 41.948 & $5/2$
& 0.972($1h_{11/2}$),\;0.026($2f_{7/2}$) \\
\hline 
 (-5, 1$\nu$1$\pi$, 1.88)& 1$\nu$ & 0.501 & 53.491 & $-3/2$
& 0.880($1h_{9/2}$),\;0.080($2f_{5/2}$)  \\
         &   1$\pi$      &  1.387      & 42.104 & $-7/2$
& 0.978($1h_{11/2}$),\;0.020($2f_{7/2}$) \\
\hline 
 (-1, 1$\nu$1$\pi$, 1.94)  &1$\nu$  &0.627  & 53.325 & $-3/2$
& 0.768($2f_{7/2}$),\;0.137($1h_{9/2},$),\;0.036($3p_{3/2}$) \\
         &    1$\pi$     &  1.319      & 41.948 & $1/2$
& 0.972($1h_{11/2},$),\;0.026($2f_{7/2}$)  \\
\hline 
 (1, 1$\nu$1$\pi$, 2.18)  &1$\nu$  &0.627  & 53.325 & $-3/2$
& 0.768($2f_{7/2}$),\;0.137($1h_{9/2},$),\;0.036($3p_{3/2}$) \\
         &    1$\pi$     &  1.556     & 42.407 & $5/2$
& 0.985($1h_{11/2},$),\;0.013($2f_{7/2}$)  \\
\hline \hline 
\end{tabular}
\label{nillson}
\end{table*}

To further elucidate the structural characteristics of the rotational bands presented in Fig.~\ref{Figure_1.pdf}, the wave function amplitudes $g_i^2$ for Bands B1, B2, B3, B4 and B5 are compiled in Table~\ref{gi2}. The quantity $g^{\sigma I}_{K\kappa}$ is defined by the expression:
\begin{equation}
g^{\sigma I}_{K\kappa} = \sum_{K',\kappa'}{\langle \Phi_{\kappa}|\hat{P}^{I}_{KK'}|\Phi_{\kappa'}\rangle^{1/2} f^{\sigma I}_{K'\kappa'}},
\end{equation}
where $g^{\sigma I}_{K\kappa}$ represents the wave function amplitudes and each index $K$ corresponds to a specific basis state within the configuration space.
In the table, the first three largest components in the wavefunction are listed. It is noted from thre table that 
at spin, $I=1^-\hbar$, the ground-state band B1 exhibits a predominant contribution from the configuration $(-1, 1\nu1\pi, 1.79)$
of about 80.0$\%$ to the total wavefunction. A second 2-qp configuration of $(-1, 1\nu1\pi, 1.81)$ has the probability of 12.0$\%$.
For the angular momentum, $I=2^-\hbar$, the dominance of the primary configuration diminishes to 65.5$\%$, while another 2-qp structure
$(-2, 1\nu1\pi, 1.87)$ emerges with considerable strength, contributing 20.3$\%$ to the wavefunction. These two 2-qp configurations continue to govern the structural evolution of band B1 up to $I=6^-\hbar$, maintaining the band's characteristic rotational behaviour.

Band B2 displays a more complex structural evolution with large mixing of the configurations of
$(-1, 1\nu1\pi, 1.81)$, $(-2, 1\nu1\pi, 1.87)$, and $(1, 1\nu1\pi, 1.87)$ at low spin values. The wavefunction composition undergoes
significant changes with increasing angular momentum, particularly as band crossing phenomena become prominent. The structural properties of band B3 reveal a dynamic competition among multiple configurations in the low-spin regime. At $I=1^-\hbar$, the band is dominated by the $(1, 1\nu1\pi, 1.87)$ configuration, which contributes 50.1$\%$ and establishes the band head. At $I=2^-\hbar$, the $(-1, 1\nu1\pi, 1.81)$ configuration becomes predominant with a 52.3$\%$ contribution, while the $(1, 1\nu1\pi, 1.81)$ configuration gains strength as the spin approaches $I=6^-\hbar$, indicating a gradual structural transition within this band. 

The positive-parity band B4 exhibits particularly intricate structural evolution. The band head at $I=1^+\hbar$ is characterized by strong dominance of the $(1, 1\nu1\pi, 2.18)$ configuration, contributing 79$\%$ to the wavefunction. However, at $I=2^+\hbar$, the band structure undergoes substantial configuration mixing, with the $(0, 1\nu1\pi, 1.82)$ configuration emerging as dominant (52.2$\%$), while the initial $(1, 1\nu1\pi, 2.18)$ configuration maintains significant
contribution (22.1$\%$). This mixing pattern suggests the onset of alignment processes that significantly modify the band's intrinsic structure. 

The positive-parity band B5 exhibits significant configuration mixing with spin. At spin~$1^+\hbar$, the dominant component $(-1,1\nu1\pi,1.94)$
contributes 63.2$\%$. For spin~$2^+\hbar$, the configuration shifts with $(1,1\nu1\pi,2.18)$ probability at 50.1$\%$ and $(0,1\nu1\pi,1.82)$
at 13.1$\%$. At higher spins ($3^+\hbar$, $4^+\hbar$), there is a significant mixing with three competing components including the configuration
($-1$,$0$ and $3$) showing substantial
probability of (32.1--33.5 $\%$). 

In Table~\ref{nillson}, the dominant spherical components in the corresponding Nilsson wavefunctions of the
quasiparticle configurations of  Table~\ref{gi2} are listed. It is noted from this table that the quasiparticle
energy of 1.79 MeV has the dominant $\nu(2f_{7/2}), \pi(2d_{5/2})$ spherical configuration. This is the
true for all the states from I=$1^-$ to $4^-$ of band B1. For I=$5^-$, the dominant
quasiparticle state of energy 1.81 MeV mostly originates from the spherical state of
$\nu(2f_{7/2}), \pi(1g_{7/2})$. For band B2 and B3, the dominant spherical configurations is
$\nu(2f_{7/2}), \pi(1g_{7/2})$.  For the  I=$1^+$  state of  band B4,  the dominant spherical configurations is $\nu(2f_{7/2}), \pi(1h_{11/2})$ whereas the higher spin states from  I=$2^+$ to $4^+$ are dominated by the    $\nu(1h_{9/2}), \pi(1h_{11/2})$ configurations. 
It can be seen from the Table~\ref{nillson} that the quasiparticle energy of 1.94 MeV corresponds to the dominant  $\nu(2f_{7/2}), \pi(1h_{11/2})$ spherical configuration for the I = $1^+$  of band B5. For the I = $2^+$ state of the band B5, the dominant spherical configuration is    $\nu(2f_{7/2}), \pi(1h_{11/2})$. For the I = $3^+$ and  $4^+$ states, the quasiparticle energy of 1.82 MeV predominantly originates from the spherical configuration  of $\nu(1h_{9/2}), \pi(1h_{11/2})$.

\subsection{Phenomenological Interpretation}
\label{pi}

The possible ground-state configuration of $^{150}$Pm may also be discussed from the Nilsson orbital systematics of the neighboring odd-$A$ nuclei. From the adopted assignments in $^{149,151}$Pm ~\cite{jones149Pm,vermeer151Pm} and $^{149}$Nd/$^{151}$Sm~\cite{budick149Nd,burke151Sm}, the relevant low-lying quasiproton orbitals near the Fermi surface are $\pi 5/2[402]$ and $\pi 7/2[404]$, while the lowest quasineutron orbital is $\nu 5/2[523]$. Accordingly, the ground-state configuration of $^{150}$Pm is expected to arise most likely from either $\pi 5/2[402]\otimes \nu 5/2[523]$ or $\pi 7/2[404]\otimes \nu 5/2[523]$. For the $\pi 5/2[402]\otimes \nu 5/2[523]$ coupling, the allowed $K^\pi$ values are $0^-$ and $5^-$. According to the Gallagher--Moszkowski rule~\cite{GM_rule}, the low-spin member is favored, suggesting a $K^\pi=0^-$ structure. In odd-odd deformed nuclei, however, the residual neutron-proton interaction may produce a Newby shift~\cite{newby}, such that the $I^\pi=1^-$ state lies below the $0^-$ state in a $K^\pi=0^-$ band. Thus, the observed ground state may naturally be associated with a $1^-$ assignment. For the alternative $\pi 7/2[404]\otimes \nu 5/2[523]$ coupling, the allowed $K^\pi$ values are $1^-$ and $6^-$. Depending on the dominant parentage of the $\nu 5/2[523]$ orbital, this configuration may also support a low-lying $1^-$ state. On the other hand, it is difficult to identify any plausible combination of low-lying quasiproton and quasineutron Nilsson orbitals that would naturally lead to a $2^-$ ground state. Therefore, the Nilsson-configuration systematics of neighboring nuclei provide additional support for assigning the ground-state spin-parity of $^{150}$Pm as $I^\pi=1^-$.

\section{Summary and Discussion}

The low lying level structure of $^{150}$Pm were populated with proton induced reaction and was proposed with relative intensities of $\gamma$ transitions and very few tentative spin parity assignment. 16 new $\gamma$ rays were placed in the level scheme resulting in 15 new levels compared to the existing data. There are distinctly two groups of levels that were placed in the level scheme. One group is confirmed to be developed on the ground state while the other may be candidates of level structure developed on some higher lying isomeric state.  

Lifetime measurement has been performed for two of the observed levels using Generalized Centroid Difference (GCD) technique utilized with Clover Ge detectors. The lieftimes of both the 128.3~keV and 177.2+X~keV levels were found to be less than 2~ns.

Some of the low-lying levels including the ground state matches well with the calculated levels from shell model, indicating the prominent role of single particle excitations in this near spherical nucleus. However, the calculation could not predict any low energy positive parity level, possibly due to restricted model space. 

Considering the deformation of the neighboring even-even $^{150}$Nd code, the quasi-particle band structures in the odd-odd $^{150}$Pm was explored using projected shell model calculation and the comparison of the results with the experimental data is quite reasonable.  

It is understood from both the calculations and the phenomenological interpretation that the ground state spin parity of $^{150}$Pm is conjectured as 1$^-$. 

In order to completely understand the structure of $^{150}$Pm, it would be further important to take up high-resolution $\gamma$ spectroscopic measurements with $^2H$ beam and with a setup that is appropriate to measure angular distribution, angular correlation and linear polarization. 

\section*{Acknowledgement}

Authors sincerely acknowledge the meaningful discussions and valuable suggestions from Prof.~M.~Saha~Sarkar and Prof.~S.~S.~Sarkar. The K-130 cyclotron operation group at VECC, Kolkata, is gratefully acknowledged for providing high quality proton beam. The INGA collaboration is acknowledged for providing the setup for the experiment.

\end{document}